\documentclass{article}

\usepackage[margin=1.2in]{geometry}
\usepackage{booktabs}
\usepackage{cite}

\usepackage{amsmath}
\usepackage{amssymb}
\usepackage{array}

\usepackage{eurosym}
\usepackage{enumitem}

\usepackage[font=normalsize]{caption} 
\usepackage{subcaption}
\usepackage[font=scriptsize]{subcaption} 
\usepackage{graphicx}
\graphicspath{{./}}
\usepackage{xcolor}
\usepackage[colorlinks=true,linkcolor=blue,citecolor=blue,urlcolor=blue]{hyperref}
\usepackage{float}
\usepackage{placeins}

\newtheorem{theorem}{Theorem}

\newtheorem{definition}[theorem]{Definition}

\input{tcilatex}

\begin{document}

\title{{\bfseries Nullclines, Subnullclines and the Asymptotic and Transient Attractors in Eco-Evolutionary Dynamics}}

\author{
\small
Krzysztof Argasinski$^{1}$\thanks{Corresponding author: \textit{argas1@wp.pl}},
Manjyot Singh Bedi$^{2}$,
Mark Broom$^{2}$\\[0.5em]
$^{1}${\small Faculty of Mathematics, Informatics, and Mechanics, University of Warsaw, ul. Banacha 202-097,}\\
{\small Warszawa, Poland}\\[0.25em]
$^{2}${\small Department of Mathematics, City St George's, University of London, EC1V 0HB, London, UK}
}

\date{}

\maketitle


\begin{abstract}
In the demographic framework, mortality payoff function describes the cost of an interaction and fertility payoff function describes its reward. So while mortality cost depends on opponent's strategy, fertility reward can be affected by the density-dependent
juvenile recruitment survival.

\medskip

This motivates an analysis of the eco-evolutionary dynamics of the classical Hawk-Dove game. It is shown that the stable and unstable equilibria (determined by the intersections of frequency and density nullclines) are connected by heteroclinic orbits, which attract nearby trajectories. The resulting bundle of trajectories leads to the discovery of the so-called subnullcines (manifolds placed
between frequency and density nullcline) before they converge to the stable
rest point. The initial isolated system is then extended by adding environmental seasonality (periodic background mortality), which acts as an external factor. This leads to complex cycling behavior and the subnullclines act as barriers to the propagation of the perturbation (resilience/resistance threshold). Thus, in a way, this paper completes, yet extends, previous works on the eco-evolutionary dynamics of games with demographic payoffs.
\end{abstract}

\section{Introduction}

The evolutionary game-theoretic framework used here builds on previous work that formulated the model in terms of demographic parameters \cite{doebeli}, particularly empirically observable birth and death rates \cite{argbr1,argbr2,argbr3}. In the new approach, growth suppression is derived from the logistic term describing juvenile recruitment survival. Consequently, the resulting eco-evolutionary dynamics are more complex. In contrast to the classical textbook theory \cite{maynard1,maynard2,hofsig1,hofsig2,BroomRychtar,Sinervo}, for a two-strategy population, the game-theoretic equilibria is governed by nullclines of the frequency dynamics (the manifold of equal growth rates for the competing strategies), while the set of ecological equilibria is shaped by the density nullcline (the manifold of zero population growth rate). The steady states of the new eco-evolutionary setup are the intersections of the frequency and density nullclines. For some parameter sets, the dynamics exhibit a stable equilibrium and a saddle point connected by a heteroclinic orbit, which contains a transient attractor for nearby trajectories \cite{Bedi2025}. In this paper, we focus on this mechanism and investigate how the geometry of the nullclines affects the stability of the rest points and the shape of the transient attracting surface.

It is worth noting that the configuration formed by the connected saddle and node is part of a larger dynamical structure. When a change in the parameters results in a saddle-node bifurcation, we observe a pattern known as a \emph{temporary ghost attractor} \cite{Hastings1,Hastings2,Morozov,Morozov2,Koch2024}. This causes trajectories to become trapped between the nullclines for some time, appearing to be in a stable state before they suddenly and rapidly switch to the actual rest point. This phenomenon is termed a \emph{long transient}. In addition, changes in parameters leading to a change in behavior are related to the concept of \emph{resilience of the system}, i.e. the ability of the system to absorb perturbation \cite{ResHolling1,ResHolling2,ResGund,ResKrak,ResMeyer,ResReed}. For example, we can imagine periodic perturbation as an inherent part of the system, similarly to the so-called flow-kick models \cite{ResMeyer,ResReed}, but in continuous time. Then, the analysis of the impact of the perturbation can be verified on the induced changes in the geometry of the nullclines. The question arises: are these issues related to each other and linked by a single mathematical concept? In this paper, we investigate this issue.

\subsection{State of the art} \label{StateOfTheArt}

The general model from previous papers \cite{argbr1,argbr2,argbr3} assumes two competing strategies, with $n_{i}$ individuals of each
strategy. The starting point is the system of the Malthusian growth
equations: 
\begin{equation}
\frac{dn_{i}}{dt}=n_{i}(t)\tau \left[ \left( e_{i}Vq^{T}(t)+\Phi \right)
\left( 1-\frac{n(t)}{K}\right) -e_{i}Dq^{T}(t)-\Psi \right] \text{ for }
i=1,2,  
\label{Eq:Malthus}
\end{equation}

where $e_1=(1,0)$ and $e_2=(0,1)$ are the canonical unit vectors in $\mathbb{R}^2$ (representing the two pure strategies), $n \; (=n_1+n_2)$ denotes the population size, $\tau=1$ is the effective interaction rate allowing for a demographic interpretation of payoffs and $V=S\bullet W$ is the survival-fertility trade-off matrix, where $\bullet$ denotes elementwise multiplication. Here, $W$ denotes the fertility matrix with entries $W_{i,j}$ describing the number of offspring produced during a game round, $D$ is the mortality payoff matrix with entries $d_{i,j}$ describing the probability of death of an $i$th-strategy player during a game round against a $j$th-strategy opponent and $S$ is the survival matrix (auxiliary notation) with entries $s_{i,j}=1-d_{i,j}$. Furthermore, $\Phi$ is the background fertility rate describing the intensity of background birth, while $\Psi$ is the background mortality rate describing the intensity of background death. $K$ is the carrying capacity, interpreted as the number of nest sites \cite{hui}, making the logistic term $\left(1-{n(t)}/{K}\right)$ represent juvenile recruitment survival, which is proportional to the fraction of free nest sites. So, every newborn checks a single random nest site and survives only if it is free, otherwise it dies. Generalizations are possible.

Demographic payoff functions $V$ and $d$ should be multiplied by effective interaction rate $\tau $ describing the intensity of the occurrence of the focal game rounds. Background mortality and fertility rates are adjusted to the timescale where $\tau =1$. In
effect, this parameter is not explicitly present in the equations. Also, note that in this structure wherein entries $V_{i,j}=s_{i,j}W_{i,j}=(1-d_{i,j})W_{i,j}$ describe the expected reproductive success of $i$th strategy survivors playing with $j$th strategy opponent, the distinction between winners and losers is not considered. Thus, it can be insufficient for a description of games with indivisible rewards. In this paper, the model will be upgraded to address this issue.

Equation~\eqref{Eq:Malthus} can be expressed in terms of replicator dynamics by applying the change of variables $q=[q_{1},1-q_{1}]$, where $q$ is the vector of strategy frequencies, $q_{1}=n_{1}/(n_{1}+n_{2})$ and $n=n_{1}+n_{2}$ denotes the total population size, as follows,
\begin{align}
\frac{dq_{1}}{dt}
= g(q,n)= q_{1}(t) &\left[\left(
e_{1}Vq^{T}(t)-q(t)Vq^{T}(t)\right)\left(1-\frac{n(t)}{K}\right)
-\left(
e_{1}Dq^{T}(t)-q(t)Dq(t)\right)\right]
\label{Eq:FrequencyRep}
\\[0.5em]
\frac{dn}{dt}
= f(q,n)= n(t) &\left[q(t)Vq^{T}(t)\left(1-\frac{n(t)}{K}\right)
-q(t)Dq(t)+\Phi\left(1-\frac{n(t)}{K}\right)-\Psi\right]
\label{Eq:PopulationRep}
\end{align}

The roots of equations~\eqref{Eq:FrequencyRep} and~\eqref{Eq:PopulationRep} determine the frequency and density nullclines, respectively. Their intersections define rest points, the stability of which is discussed in detail later on. The description of notation and symbols can be referred to from Table~\ref{tab:ImportantSymbols}.

\begin{table}[t!]
\centering
\vspace{-1em}
\setlength{\tabcolsep}{6pt}
\renewcommand{\arraystretch}{1.15}
\begin{tabular}{@{}p{0.25\textwidth}p{0.74\textwidth}@{}}
\toprule
\textbf{Symbol} & \textbf{Description} \\
\midrule
$n_i$ & number of individuals using strategy \(i\) \\

$e_i$ & unit vector \\

$n=n_1+n_2$ & population size \\

$q_i=n_i/(n_1+n_2)$ & frequency of the \(i\)-th strategy \\

$\tau=1$ & effective interaction rate allowing for a demographic interpretation of payoffs in the differential equations \\

$W$ & fertility matrix, with entries \(W_{i,j}\) describing the number of offspring produced during a game round \\

$D$ & mortality payoff matrix, with entries \(d_{i,j}\) describing the probability of death of an \(i\)-strategy player during a game round against a \(j\)-strategy opponent \\

$S$ & survival matrix, with entries \(s_{i,j}=1-d_{i,j}\) \\

$V=S\bullet W$ & survival--fertility trade-off matrix, where \(\bullet\) denotes elementwise multiplication \\

$\Phi$ & background fertility rate, interpreted as the intensity of background birth \\

$\Psi$ & background mortality rate, interpreted as the intensity of background death \\

$\left(1-\dfrac{n(t)}{K}\right)$ & logistic juvenile recruitment survival, proportional to the fraction of free nest sites \\

$K$ & carrying capacity, interpreted as the number of nest sites \\

$g(q,n)$ &  rate of increase in the frequency of the first strategy \\

$f(q,n)$ &  rate of increase in the population density \\

$g_q=\partial g/\partial q$, \(g_n=\partial g/\partial n\) & partial derivatives of \(g(q,n)\) \\

$f_q=\partial f/\partial q$, \(f_n=\partial f/\partial n\) & partial derivatives of \(f(q,n)\) \\

$C$ & win-probability matrix, with entries \(p_{i,j}\) \\

$\mathcal{S}^{\varepsilon}_q(q_{d})$ & frequency subnullcline \\

$\mathcal{S}^{\varepsilon}_n(q_{d})$ & density subnullcline \\

$d$ & Hawk vs. Hawk probability of death during a game round \\

$d^w$, \(d^l\) & death probabilities associated with winner and loser roles, respectively; simplified to \(d\) in the replicator equations for the Maynard Smith model \\

\bottomrule
\end{tabular}
\caption{Important symbols}
\label{tab:ImportantSymbols}
\vspace{-1em}
\end{table}

\subsection{Plan of the paper}

In this paper, we complete the demographic game framework by introducing three theoretical tools, constituting the following research goals:
\begin{enumerate}[label=\alph*)]
    \item Clarification of the stability conditions based on nullcline intersections.

    \item We know from previous papers that the eco-evolutionary dynamics converge to a stable rest point along a heteroclinic orbit connecting the stable and unstable rest points \cite{Bedi2025}. The manifold constituted by this orbit is contained in the area bounded by the intersecting frequency and density nullclines. Here, we propose the construction of an approximate attractor of the \emph{bundle of trajectories} based on a generalization of the concept of the isocline.

    \item The approximation of the attracting surface can also be useful to approximate the so-called ghost attractors and for the analysis of resilience against external perturbations.
\end{enumerate}

Real-world populations are rarely entirely isolated systems, such as bacterial colonies in a petri dish. Rather, they are parts of larger and more complex ecosystems affected by other external, often periodic, environmental factors. We show that eco-evolutionary systems based on juvenile recruitment survival, which affects fertility reward, are sensitive to external factors acting in the background, such as seasonal mortality. In particular, mixed polymorphic states may be sensitive to environmental factors that are not directly linked to the game. This may affect the shape of trajectories or even destabilize the stable rest point.

\section{Analytical results}

The Jacobian matrix is composed of the gradients responsible for isolating the flow for each state variable into its partial derivatives with other respective dynamic variables. For our two-strategy density-dependent replicator system \eqref{Eq:FrequencyRep} and~\eqref{Eq:PopulationRep}, we get

\renewcommand{\arraystretch}{1.5} 
\begin{equation*}
 J = \begin{bmatrix}            
    \frac{\partial g}{\partial q_1} & \frac{\partial g}{\partial n}\\
     \frac{\partial f}{\partial q_1} &\frac{\partial f}{\partial n} \\
\end{bmatrix}
= \begin{bmatrix}
    g_{q} & g_{n}\\
    f_{q} & f_{n} \\
\end{bmatrix}.
\label{eq:Jacobian}
\end{equation*}

The local behaviour around a rest point is determined by the eigenvalues of
the Jacobian matrix. The characteristic polynomial of which is,
\begin{equation}
\lambda^2-(g_q+f_n)\lambda
+
(g_qf_n-g_nf_q).
\label{Eq:CharacteristicPolynomial}
\end{equation}

As established in Theorem~2 of~\cite{argbr3}, an intersection $(\hat{n},\hat{q})$ is stable if the following are satisfied,
\begin{enumerate}[label=\alph*),ref=\alph*)]
    \item \label{cond:a}
    \begin{equation*}
    g_{q}(\hat{n},\hat{q})<\left\vert f_{n}(\hat{n},\hat{q})\right\vert
    \label{Eq:Condition(a)}
    \end{equation*}

    \item \label{cond:b}
    \begin{equation*}
    \dfrac{dg(\tilde{n}(q),q)}{dq_{1}}
    =g_{q}(\hat{n},\hat{q})-g_{n}(\hat{n},\hat{q})
    \dfrac{f_{q}(\hat{n},\hat{q})}{f_{n}(\hat{n},\hat{q})}<0
    \label{Eq:Condition(b)}
    \end{equation*}
\end{enumerate}

Condition~\ref{cond:a} shows that juvenile recruitment survival can stabilize an equilibrium that is unstable in the classical unsuppressed replicator dynamics (as illustrated in Example~1 of~\cite{argbr3}). Condition~\ref{cond:b} concerns the derivative of the frequency equation~\eqref{Eq:FrequencyRep}. It holds for \(g_{n}<0\) (\(g_{n}>0\)) when the slope of the density nullcline along the \(q_{1}\)-axis is greater than (is smaller than) the corresponding slope of the frequency nullcline. This confirms that the geometry of the nullclines underlies the eco-evolutionary feedback mechanism that determines the stability of the rest points. We examine both conditions through a geometric lens in the following subsection.

\subsection{Geometric stability conditions}
\begin{minipage}[htbp]{0.5\textwidth}
     Assume the general case from Figure~\ref{AnglesForInequality} where $A$ and $B$ are the vertical and horizontal axes, respectively, representing general variables in a dynamical system. With \(A_A=\partial \dot A/\partial A\) and
\(A_B=\partial \dot A/\partial B\), the gradients of the two right-hand
sides are therefore
\[
\nabla A=(A_A,A_B),\qquad \nabla B=(B_A,B_B).
\] 

The slope of the gradient \(\nabla A\), measured with respect to the \(A\)-axis,
is
\[
S_g(\nabla A)=\tan(\alpha_A)=\frac{A_B}{A_A}.
\]

The \(A\)-nullcline is locally tangent to any vector \((a,b)\) satisfying
\[
\nabla A\cdot(a,b)=aA_A+bA_B=0.
\]
Hence
\[
b=-a\frac{A_A}{A_B}.
\]
    
\end{minipage}
\hfill
\begin{minipage}[htbp]{0.5\textwidth}
    \centering
    \includegraphics[width=0.8\linewidth]{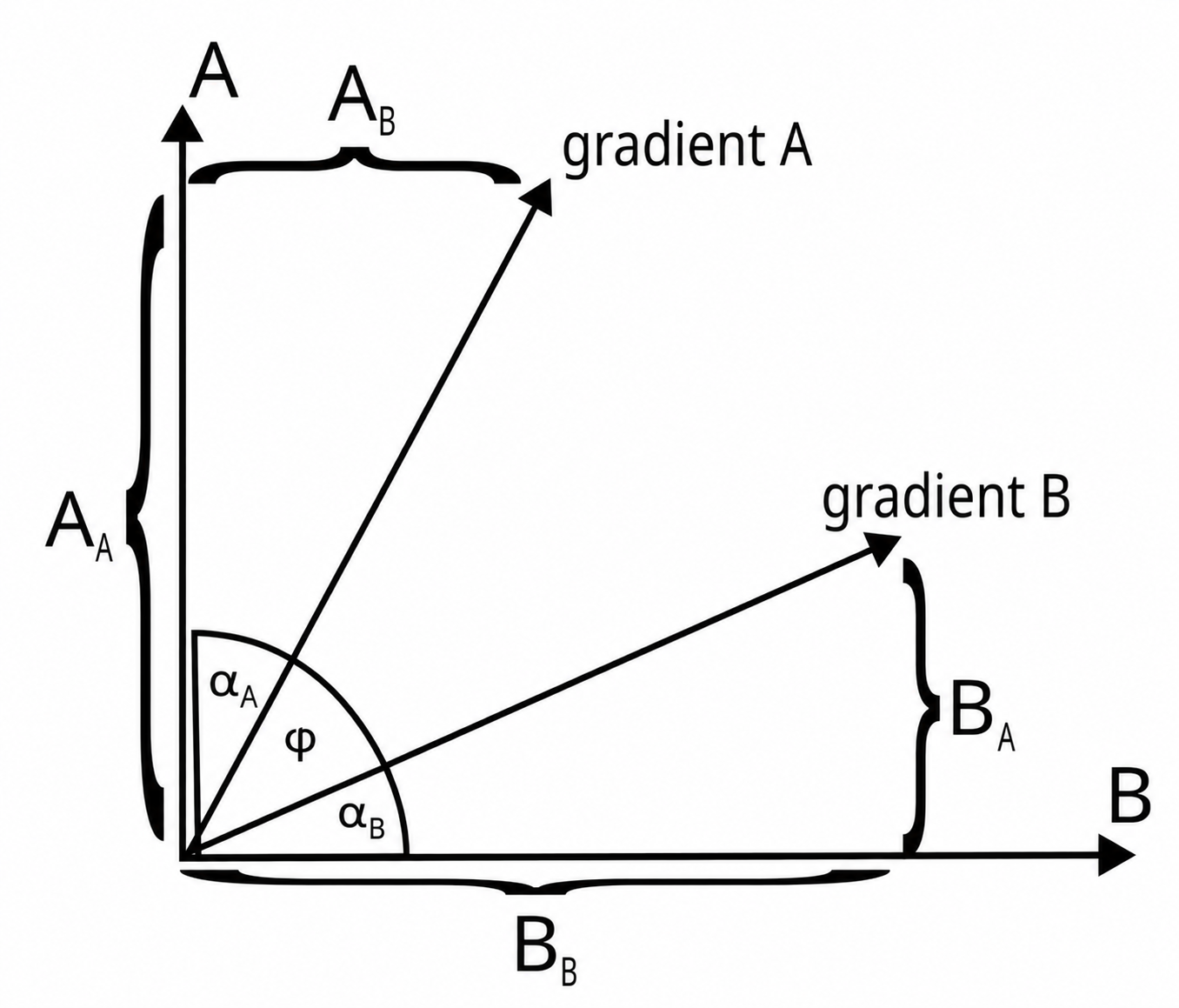}
    \captionof{figure}{Gradients in the plane spanned by ($A$) and ($B$), together with their angles relative to the corresponding focal axes.}
    \label{AnglesForInequality}
\vspace{1em}
\end{minipage} 

Thus, if \(S_A^A\) denotes the slope of the \(A\)-nullcline (described by superscript) measured with
respect to the \(A\)-axis (described by subscript) and \(S_B^A\) denotes the same nullcline slope
measured with respect to the \(B\)-axis, then
\[
S_A^A=\frac{b}{a}=-\frac{A_A}{A_B},
\qquad
S_B^A=\frac{a}{b}=-\frac{A_B}{A_A}.
\]

Since the gradient is orthogonal to the nullcline, its slope can be expressed in terms of the nullcline slope,
\[
S_A^A=-\frac{1}{S_g(\nabla A)},
\qquad
S_B^A=-S_g(\nabla A).
\]
Therefore, 
\[
S_A^A=\frac{1}{S_B^A}.
\]

Similarly, for the \(B\)-nullcline,
\[
S_A^B=-\frac{B_A}{B_B},
\qquad
S_B^B=-\frac{B_B}{B_A}.
\]

The trace (equivalent to condition \ref{cond:a} is,
\[
A_A+B_B<0.
\]

The determinant, which is the cross product of both gradients (equivalent to condition \ref{cond:b} is,
\[
\begin{aligned}
A_AB_B-A_BB_A>0
&\iff
A_AB_B
\left(
1-\frac{A_BB_A}{A_AB_B}
\right)>0 \\[0.5em]
&\iff
A_AB_B
\left(
1-S_B^A S_A^B
\right)>0 \\[0.5em]
&\iff
A_AB_B
\left(
1-\frac{S_A^B}{S_A^A}
\right)>0 \\[0.5em]
&\iff
A_AB_B
\left(
1-\frac{S_B^A}{S_B^B}
\right)>0 .
\label{Eq:SlopeBasedDeterminant}
\end{aligned}
\]

Thus, if both focal directions are locally stable, \(A_A<0\) and \(B_B<0\),
then the trace condition is automatically satisfied and \(A_AB_B>0\).
In this case, the determinant condition reduces to
\[
1-\frac{S_B^A}{S_B^B}>0.
\]

If \(S_B^A\) and \(S_B^B\) have opposite signs, the inequality holds
trivially. If they have the same sign, then the condition is equivalently
\[
|S_B^A|<|S_B^B|.
\]
That is, measured with respect to the \(B\)-axis, the \(A\)-nullcline must
be locally flatter than the \(B\)-nullcline. If instead one focal direction is locally stable and the other is locally unstable, so that \(A_AB_B<0\), then for the determinant to be positive
\[
1-\frac{S_B^A}{S_B^B}<0,
\]
or equivalently, when the slopes have the same sign,
\[
|S_B^A|>|S_B^B|.
\]
In this case, the trace condition \(A_A+B_B<0\) must still be checked
separately. \medskip

Therefore, local stability can be inferred geometrically through the
relative ordering of the two nullclines in a neighbourhood of the
intersection i.e. by determining which nullcline lies locally above the
other. In the non-degenerate case this ordering is captured by the linear
terms and hence by the relative slopes at the intersection. \medskip

In the degenerate cases where the two
nullclines have the same tangent at the rest point, linearisation is
inconclusive. The geometric method is useful as the relevant ordering is no longer determined
by the linear terms, but by the first non-vanishing higher-order terms in
the local expansions of the nullclines. Stability may then be inferred from
this higher-order nullcline ordering, together with the corresponding local
direction of flow. \medskip

When two rest points are connected by monotonous part of an attracting
nullcline, then the direction of the flow along that segment cannot change
unless the segment intersects the nullcline associated with the other
variable. Indeed, along a given nullcline one component of the vector field
vanishes identically, so the direction of motion along it is determined by
the sign of the other component. This sign can change only where the second
component also vanishes, namely at an intersection with the other nullcline. Building on this geometric intuition, we need
\begin{enumerate}
\item Decaying Perturbations: Trace, $tr(J)<0$, i.e.

\begin{equation}
g_{q}+f_{n}<0.  \label{eq:Trace}
\end{equation}

\vspace{0.5em} The trace corresponds to the sum of the eigenvalues of the system, which in turn determine the local behaviour of the system near an equilibrium point. A negative trace implies that trajectories converge to the equilibrium, ensuring stability. A positive trace indicates divergence and instability, while a zero trace suggests possible neutral stability, oscillations or more complex dynamics that require further analysis.

\begin{figure}[htbp]
\centering
\includegraphics[width=0.75\linewidth]{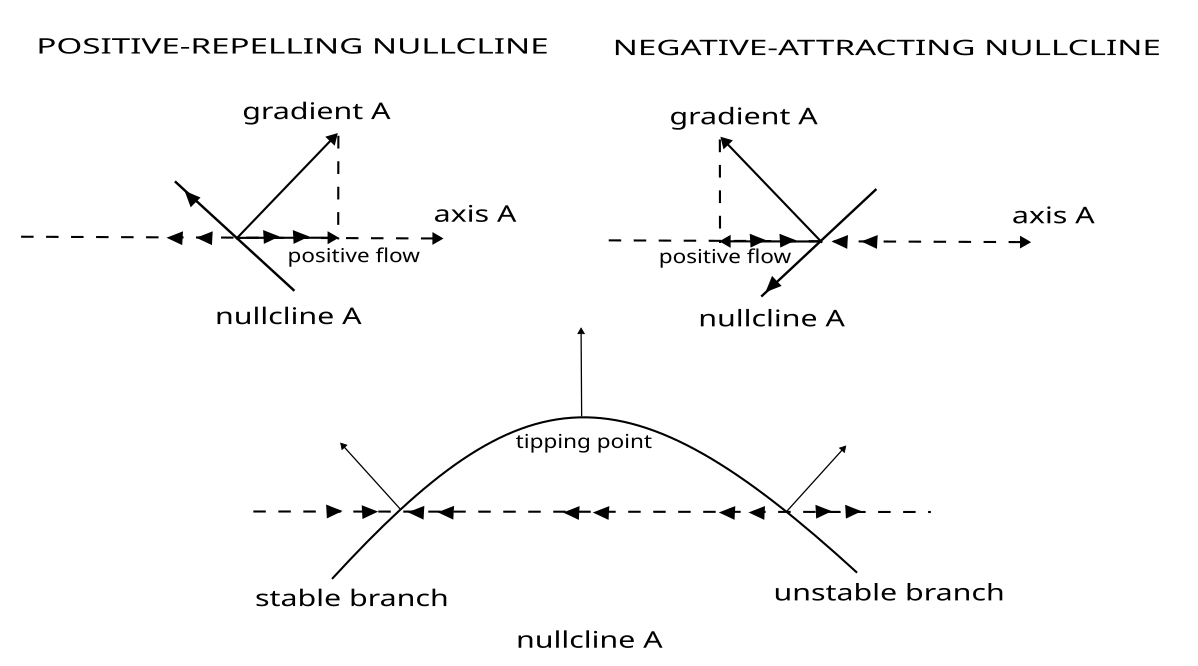}
\captionof{figure}{Stability/instability of the nullclines results from the sign of
the trace elements of the Jacobian. Figure shows how the geometry of the
nullclines changes for the same counterdiagonal entries of the Jacobian.}
\label{NullStab}
\end{figure}

\item Saddle Exclusion: Determinant, $det(J)>0$, i.e.

\begin{equation}
\begin{vmatrix}
g_{q} & g_{n} \\ 
f_{q} & f_{n}%
\end{vmatrix}%
>0.  \label{eq:2dcondition}
\end{equation}

In vector product representation, between gradients of the frequency and density equations,
\begin{equation}
\big(\mathbf{q}_{1}\times \mathbf{n}\big)>0.  \label{eq:2dconditionVec}
\end{equation}

Geometric interpretation, in terms of the lengths of the gradients and the angle $\phi$ between them leads to,
\begin{equation*}
\Vert \mathbf{q_1}\Vert \Vert \mathbf{n}\Vert \sin \phi >0.
\end{equation*}

\begin{figure}[htbp]
\centering
\includegraphics[width=0.9\linewidth]{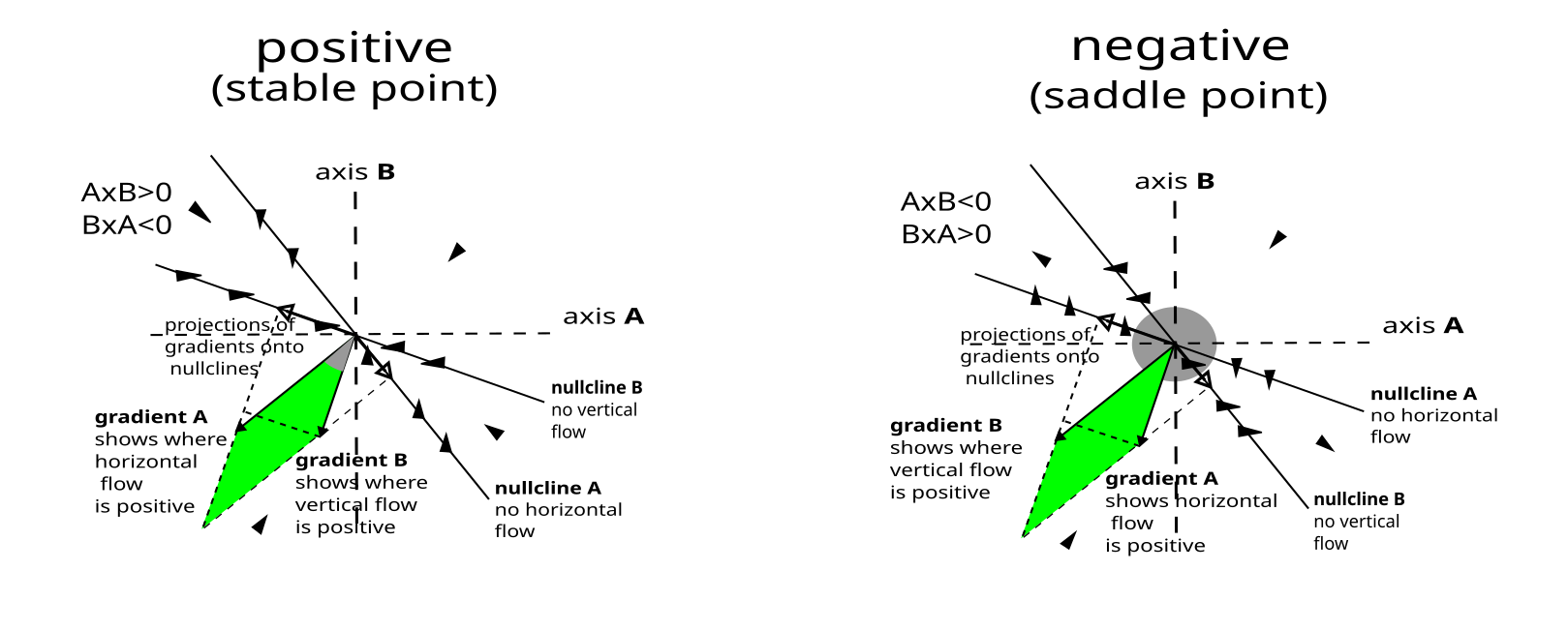}
\caption{Intuitive explanation of the relationship between stability of the
rest points and the geometry of the nulllcines.}
\label{CrossProduct}
\end{figure}

Algebraically (and analogously to the general case from \eqref{Eq:SlopeBasedDeterminant}), it has the form 
\[
\begin{aligned}
g_{q}f_{n}-g_{n}f_{q} > 0
&\;\Rightarrow\;
g_{q}f_{n}
\left(
1-\frac{g_{n}f_{q}}{g_{q}f_{n}}
\right)>0 \\[0.5em]
&\;\Longleftrightarrow\;
g_{q}f_{n}
\left(
1-S_n^q S_q^n
\right)>0 \\[0.5em]
&\;\Longleftrightarrow\;
g_{q}f_{n}
\left(
1-\frac{S_q^n}{S_q^q}
\right)>0 \\[0.5em]
&\;\Longleftrightarrow\;
g_{q}f_{n}
\left(
1-\frac{S_n^q}{S_n^n}
\right)>0 
\end{aligned}
\]

\item Local Phase Portrait: Discriminant \(\Delta\) and $Curl(F)$ i.e.

The discriminant determines whether the eigenvalues are real or complex,
which is responsible for distinguishing node-like behaviour from spiral-like behaviour by highlighting the boundary of where the flow changes from direct/straight to rotational. Curl, on the other hand measures the local rotational tendency of the vector field.
\[
\begin{aligned}
\Delta=b^{2}-4ac
&= \left(g_q+f_n\right)^2
   -4\left(g_q f_n-g_n f_q\right) \\
&= g_q f_n
   \left[
      \frac{g_q^2+2g_q f_n+f_n^2}{g_q f_n}
      -4\left(
          1-\frac{g_n f_q}{g_q f_n}
        \right)
   \right] \\
&= g_q f_n
   \left[
      \frac{g_q}{f_n}
      +2
      +\frac{f_n}{g_q}
      -4
      +4\frac{g_n f_q}{g_q f_n}
   \right] \\
&= g_q f_n
   \left[
      \frac{g_q}{f_n}
      +\frac{f_n}{g_q}
      +4\frac{g_n f_q}{g_q f_n}
      -2
   \right].
\end{aligned}
\]

If both focal directions are locally stable, then for the discriminant to be positive (negative) and yield node-like behavior (spiral-like behavior), the following inequility must hold;
\[
\left( \dfrac{g_{q}}{f_{n}}+\dfrac{f_{n}}{g_{q}}-2\right) +4\dfrac{S_{n}^{q}
}{S_{n}^{n}} >(<)0 
\]

Now, we shift our attention to determine the direction of the circular component using,
\begin{equation*}
\begin{aligned}
\operatorname{Curl}(\mathbf{F})
&= g_{n}-f_{q} \\
&= g_{q}\frac{g_{n}}{g_{q}}-f_{n}\frac{f_{q}}{f_{n}} \\
&= g_{q}(-S_n^q)-f_{n}(-S_q^n) 
\end{aligned}
\end{equation*}

Clockwise local rotation occurs when the curl is negative, hence
\begin{equation*}
\begin{aligned}
 g_{q}(-S_n^q)-f_{n}(-S_q^n) &<0 \\
 g_{q}(-S_n^q) &< f_{n}(-S_q^n) \\
 g_{q}S_n^q &> f_{n}S_q^n 
\end{aligned}
\end{equation*}
This leads to the following condition expressed along the same axis

\begin{equation*}
 g_{q}S_n^q > \frac{f_{n}}{S_n^n} .
\end{equation*}

Thus, together with the trace and the determinant, these conditions constitute the complete stability framework, including spirals. 
\end{enumerate}

Contrary to the common intuition that ecological dynamics should be ``faster'' than selection dynamics, trajectories do not necessarily converge to a small neighborhood of the density nullcline. Instead, they enter the region bounded by the two nullclines and may, for example, be attracted to a neighborhood of the heteroclinic orbit connecting their intersections. This heteroclinic connection can lie between the nullclines while remaining at some distance from both. Since nullclines are not invariant objects, but merely surfaces along which the flow is locally ``horizontal'' or ``vertical,'' they cannot themselves serve as attractors for the trajectories. We therefore require a different object, one closely related to both nullclines that can capture the attracting structure of the dynamics. In the next subsection, we propose such an object.

\subsection{Refining the stability conditions through relative isoclines}

We know that on one nullcline vector field is "horizontal"\ while on the other, it is "vertical". Inside the area limited by nullclines, the slope of the flow monotonously changes from "horizontal" to "vertical". Therefore,
there exists a point where the slope of the flow equals the slope of the particular nullcline. Below this point, the dynamics are repelled from the nullcline, while from above, they are attracted to it. This intuition is
formalized by the concept of relative isoclines.

\begin{definition}
For the vector field given by equations $\dot{x}=f(x,y)$ , $\dot{y}=g(x,y)$,
and manifold $m$ defined by function $y=m(x)$, isocline relative to $m$ is
the manifold $w$ where the slope of the dynamics is equal to the slope of
the manifold $m$. Then on the manifold$\ w$ the following condition is
satisfied: 
\begin{equation}
\frac{g(x,y)}{f(x,y)}=\frac{dm(x)}{dx}  \label{rel-iso}
\end{equation}
\end{definition}

\begin{figure}[!htbp]
\centering
\includegraphics[width=0.6\linewidth]{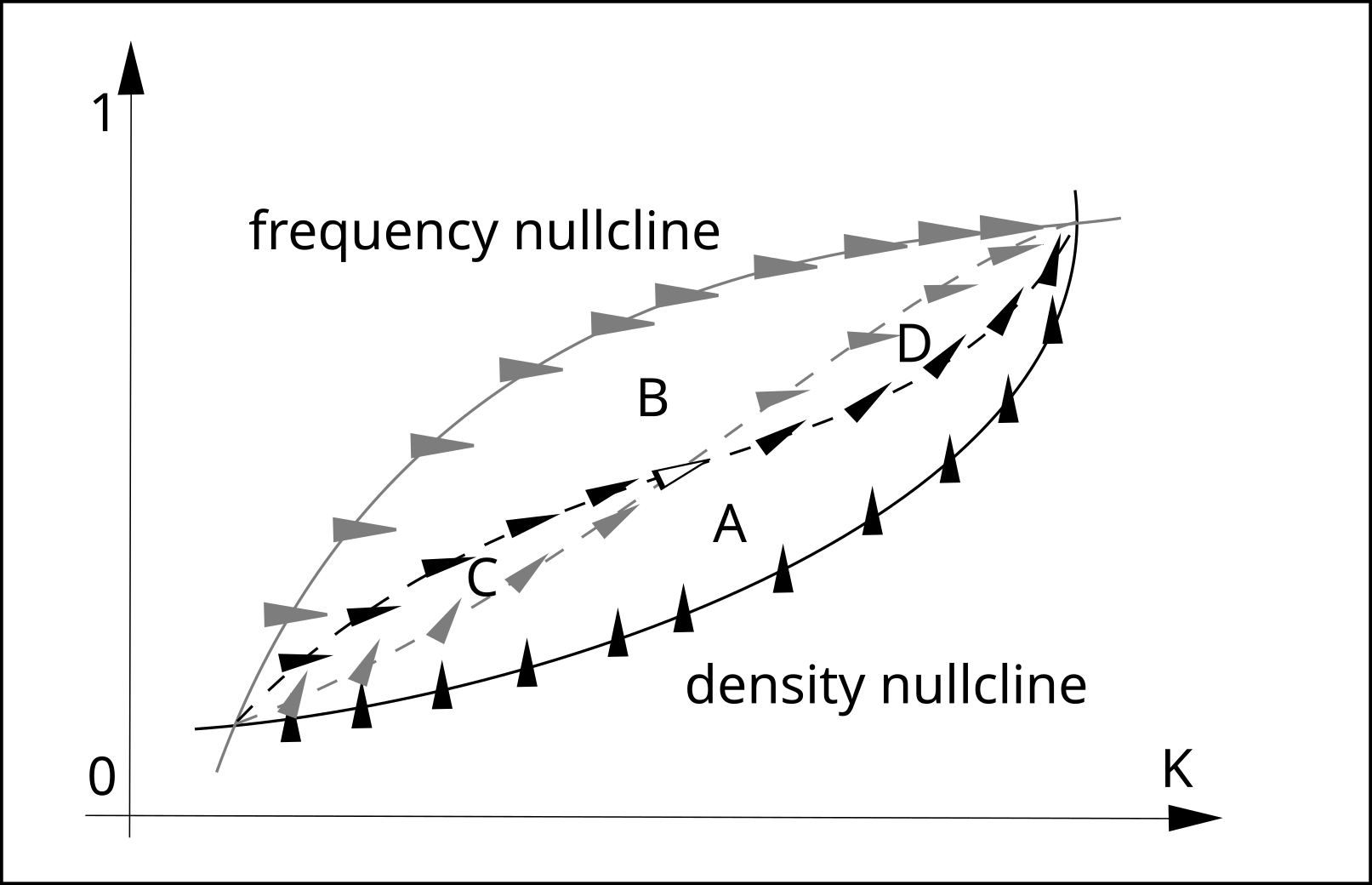}
\caption{Schematic presentation of subnullclines: A is the region of
repellence from the density nullcline, B is the region of repellence from the
frequency nullcline, C is the region of repellence from both nullclines
(regions of repellence overlap when nullclines move away from each other) and
D is the region of attraction to both nullclines (regions of attraction
overlap when nullclines approach each other). Both subnullclines intersect
at the point where the two nullclines are parallel.}
\label{Figure:Subnullclines}
\end{figure}

Relative isoclines for frequency and density nullclines, hereafter referred to as \emph{subnullclines}, constitute borders of regions of repellency and attraction of the trajectories. Both subnullclines intersect at the point where the nullclines are parallel, as shown in Figure~\ref{Figure:Subnullclines}.

We can expect that, for a range of initial conditions scattered across the phase space, the trajectories will come very close and form \emph{bundles} which then converge to the neighborhood limited by subnullclines. This region is a subset of the area bounded by the nullclines but much smaller.

\section{Example: The classical John Maynard Smith Hawk-Dove game}

In matrix form, the corresponding fertility and mortality functions from Section~\ref{StateOfTheArt} can be written as,
\begin{equation*}
V=
\left(
\begin{array}{c|cc}
 & H & D \\ \hline
H & 0.5F & F \\
D & 0 & 0.5F
\end{array}
\right),
\qquad
D=
\left(
\begin{array}{c|cc}
 & H & D \\ \hline
H & 0.5d & 0 \\
D & 0 & 0
\end{array}
\right).
\end{equation*}
Accordingly, the matrix \(V-D\) is precisely the classical Hawk-Dove payoff matrix \cite{maynard1,maynard2}, which has been studied and examined quite widely in the framework of replicator dynamics,
\begin{equation*}
\left( 
\begin{array}{c|cc}
& H & D \\ \hline
H & 0.5\left( F-d\right)  & F \\ 
D & 0 & 0.5F%
\end{array}%
\right) .
\end{equation*}

As the two strategy frequencies are complementary, either can be used as the focal frequency variable. We choose the Dove frequency, which simplifies calculations since Dove vs Dove contests do not incur any cost. The payoff functions, written in terms of the Dove frequency, are as follows:

\[
\begin{array}{@{}lcc@{}}
\toprule
& \text{Fertility payoff} & \text{Mortality payoff} \\
\midrule
\text{Hawk} 
& V_{h} = 0.5\left(1+q_{d}\right)F 
& D_{h} = q_{h}0.5d \\[0.3em]

\text{Dove} 
& V_{d} = 0.5q_{d}F 
& D_{d} = 0 \\[0.3em]

\text{Average} 
& \bar{V} = 0.5F 
& \bar{D} = q_{h}^{2}0.5d \\
\bottomrule
\end{array}
\]

\subsection{Nullcline structure of the frequency and density dynamics}
Since Dove mortality is zero, the replicator dynamics for the Dove frequency simplify considerably. Hence, for \(q_{h}=1-q_{d}\), the system can be written as,
\begin{align}
\frac{dq_{d}}{dt} = g(q,n) &= 0.5q_{d}(t)\Big( 1-q_{d}(t)\Big) \left[ \Big(
1-q_{d}(t)\Big) d-\left( 1-\frac{n(t)}{K}\right) F\right], 
\label{repq}
\\[0.5em]
\frac{dn}{dt} = f(q,n) &= n(t)\left[ \Big( 0.5F+\Phi \Big) \left( 1-\frac{n(t)}{K}
\right) -\Big( \big( 1-q_{d}(t)\big) ^{2}0.5d+\Psi \Big) \right].
\label{repn}
\end{align}
\begin{flushright}
\textit{For proof see Appendix~\ref{Appendix 1}}
\end{flushright}

Frequency and density nullclines are,
\begin{gather}
\tilde{q}_{d}(n)=1-\left( 1-\frac{n}{K}\right) \frac{F}{d}
\label{qnullcline} \\[0.5em]
\tilde{n}(q_{d})=\left[ 1-\frac{\left( 1-q_{d}\right) ^{2}0.5d+\Psi }{
0.5F+\Phi }\right] K,  \label{nnullcline}
\end{gather}

The intersections governed by \eqref{qnullcline} and \eqref{nnullcline} constitute the two rest points as follows,
\begin{align*}
\check{q}_{d}
&= 0.5-\frac{\Phi-\sqrt{\left(\Phi-0.5F\right)^{2}-2F\left(\frac{F\Psi}{d}-\Phi\right)}}{F}, \\
\check{n}
&= \left[1-\frac{\left(0.5+\dfrac{\Phi-\sqrt{\left(\Phi-0.5F\right)^{2}-2F\left(\frac{F\Psi}{d}-\Phi\right)}}{F}\right)^{2}0.5d+\Psi}{0.5F+\Phi}\right]K, \\
\hat{q}_{d}
&= 0.5-\frac{\Phi+\sqrt{\left(\Phi-0.5F\right)^{2}-2F\left(\frac{F\Psi}{d}-\Phi\right)}}{F}, \\
\hat{n}
&= \left[1-\frac{\left(0.5+\dfrac{\Phi+\sqrt{\left(\Phi-0.5F\right)^{2}-2F\left(\frac{F\Psi}{d}-\Phi\right)}}{F}\right)^{2}0.5d+\Psi}{0.5F+\Phi}\right]K.
\end{align*}
\begin{flushright}
\textit{For proof see Appendix~\ref{Appendix 2}}
\end{flushright}

The stability of the rest points is algebraically determined by conditions~\ref{cond:a} and \ref{cond:b}. In the case of a stable frequency nullcline, condition~\ref{cond:a} is always satisfied and condition~\ref{cond:b} need not be evaluated explicitly. At the intersection, the right-hand side of the frequency equation~\eqref{Eq:FrequencyRep} vanishes, that is, \(g(q,n)=0\), whereas condition~\ref{cond:b} determines the direction in which the value and consequently the sign, of \(g(q,n)\) changes. This sign is negative for stable rest points and positive for unstable ones.

Moreover, the sign of the right-hand side of the frequency dynamics is preserved along the density nullcline between the two intersections and can change only on the frequency nullcline. Hence, such a sign change must occur at an intersection with the density nullcline. It follows that the stability of a given intersection can be inferred from the sign of \(g(q,n)\) evaluated on the stable density nullcline. Therefore, the intersection \((\hat{q}_{d},\hat{n})\) is stable, whereas the intersection \((\check{q}_{d},\check{n})\) is unstable (see Appendix~\ref{Appendix 2}).

This, however, does not imply that trajectories converge to the density nullcline itself. As shown in earlier studies~\cite{argbr1,argbr2,argbr3}, trajectories of the eco-evolutionary system converge to a neighborhood of the heteroclinic orbit connecting the nullcline intersections. Since an analytically explicit expression for this orbit is difficult to obtain, we instead seek an object that captures its attracting properties, together with those of the neighboring trajectories. This construction is developed in the next subsection.

\subsection{Subnullcline structure of the coupled eco-evolutionary dynamics}

Now, we can complete the analytical results by calculating the subnullclines, which attract the
trajectories before they reach the stable rest point. The frequency subnullcline is the manifold where the following
equation is satisfied:
\begin{equation}
\frac{g(x,y)}{f(x,y)}=\dfrac{F}{dK}.
\end{equation}

Since the derivative of the frequency nullcline (\ref{qnullcline}) equals $
\dfrac{F}{dK}$ , this manifold is described by $\mathcal{S}^{\varepsilon}_q = \{(q,n): |\dot{q}| < \varepsilon_q \}$
\begin{equation}
n=\mathcal{S}^{\varepsilon}_q(q_{d})=\frac{-B_{q}+\sqrt{B_{q}^{2}-4A_{q}C_{q}}}{2A_{q}}
\end{equation}
where 
\begin{align*}
A_{q} &= -\left( 0.5F + \Phi \right)\dfrac{F}{dK^{2}}, \\
B_{q} &= \left[
  \frac{
    0.5F + \Phi - \left( 1 - 2q_{d}(t) + q_{d}(t)^{2} \right) 0.5d - \Psi
  }{d}
  - \left( q_{d}(t) - q_{d}^{2}(t) \right) 0.5
\right] \dfrac{F}{K}, \\
C_{q} &= 0.5\left[
  \left( q_{d}(t) - q_{d}^{2}(t) \right) F
  - \left( q_{d}(t) - 2q_{d}^{2}(t) + q_{d}^{3}(t) \right) d
\right].
\end{align*}
\begin{flushright}
\textit{For proof see Appendix~\ref{Appendix 3}}
\end{flushright}

Similarly, the density subnullcline is $\mathcal{S}^{\varepsilon}_n = \{(q,n): |\dot{n}| < \varepsilon_n \}$, 
\begin{equation}
n=\mathcal{S}^{\varepsilon}_n(q_{d})=\frac{-B_{n}+\sqrt{B_{n}^{2}-4A_{n}C_{n}}}{2A_{n}}
\end{equation}
where 
\begin{align*}
A_{n} &= \dfrac{\left( 0.5F + \Phi \right)^{2}}{K}, \\
B_{n} &= -\left[
  \left( 0.5F + \Phi \right)^{2}
  - \left( 0.5F + \Phi \right)
    \left[ \left( 1 - q_{d}(t) \right)^{2} 0.5d + \Psi \right]
  - 0.5 q_{d}(t)\left( 1 - q_{d}(t) \right)^{2} Fd
\right], \\
C_{n} &= 0.5 q_{d}(t)\left( 1 - q_{d}(t) \right)^{2} dK
\left[ \left( 1 - q_{d}(t) \right)d - F \right].
\end{align*}
\begin{flushright}
\textit{For proof see Appendix~\ref{Appendix 4}}
\end{flushright}

These subnullclines intersect at the point where the slopes of the frequency and density nullclines are equal,
\begin{equation}
\dfrac{F}{dK}=\dfrac{0.5F+\Phi }{\left( 1-q_{d}(t)\right) dK},
\end{equation}
which leads to 
\begin{equation}
\breve{q}_{d}(t)=0.5-\frac{\Phi }{F}.  \label{SubInter}
\end{equation}

This provides a good approximation of the narrowest point in the channel between the nullclines.

\section{Numerical simulations}
\begin{figure}[htbp]
\centering
\begin{subfigure}[t]{0.7\textwidth}
    \centering
    \includegraphics[width=\linewidth]{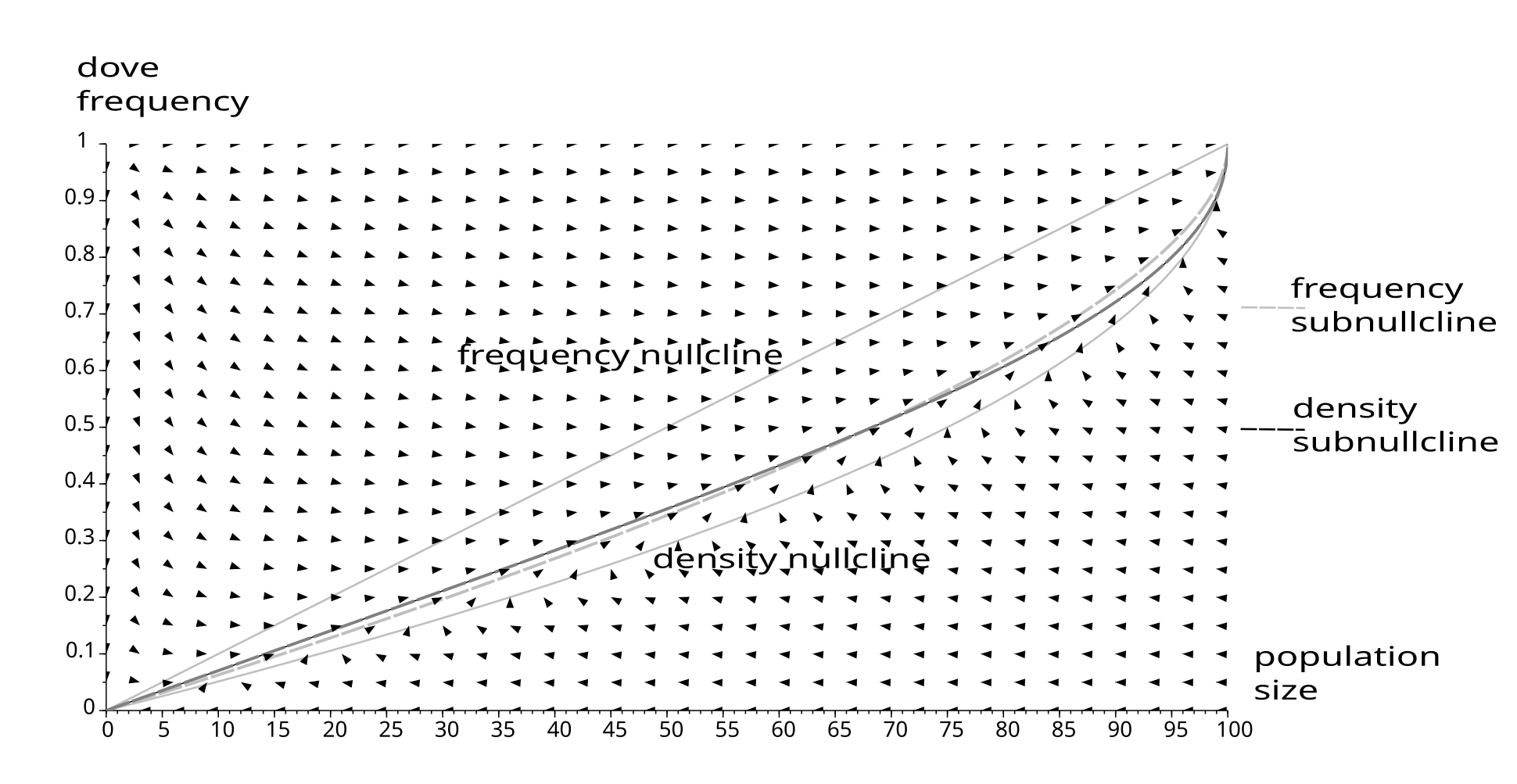}
    \caption{Slope field and the structure of nullclines and subnullclines}
    \label{fig:SlopeField}
\end{subfigure}
\begin{subfigure}[t]{0.7\textwidth}
    \centering
    \includegraphics[width=\linewidth]{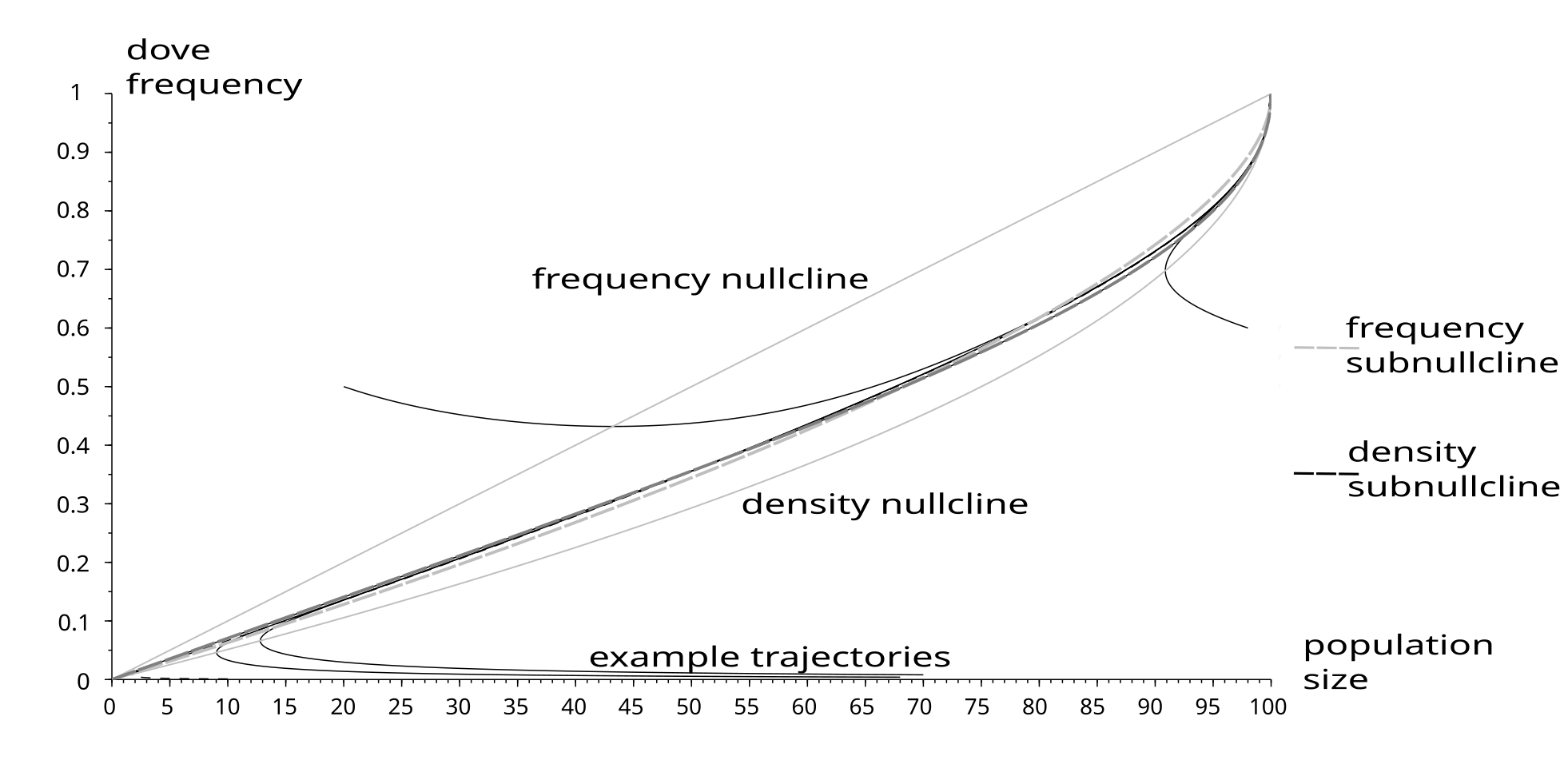}
    \caption{Example trajectories: in the real-life example, regions bounded by subnullclines are very narrow. All trajectories converge to a close neighbourhood of the subnullclines.}
    \label{fig:ExampleTrajectories}
\end{subfigure}

\caption{Dynamics of the eco-evolutionary system with model parameters: $F=0.5$, $d=0.5$, $\Phi=0$, $\Psi=0$, and $K=100$.}
\label{fig:main}
\end{figure}

The dynamics of the eco-evolutionary system (\ref{repq}, \ref{repn}) are illustrated in Figure~\ref{fig:main}, wherein panel~\ref{fig:SlopeField} displays the slope field together with the nullclines and their corresponding subnullclines. When trajectories are added, panel~\ref{fig:ExampleTrajectories} shows that they first converge toward the neighborhood of the subnullclines before turning toward the stable rest point.

\begin{figure}[htbp!]
\centering
\begin{subfigure}[t]{0.49\textwidth}
    \centering
    \includegraphics[width=\linewidth]{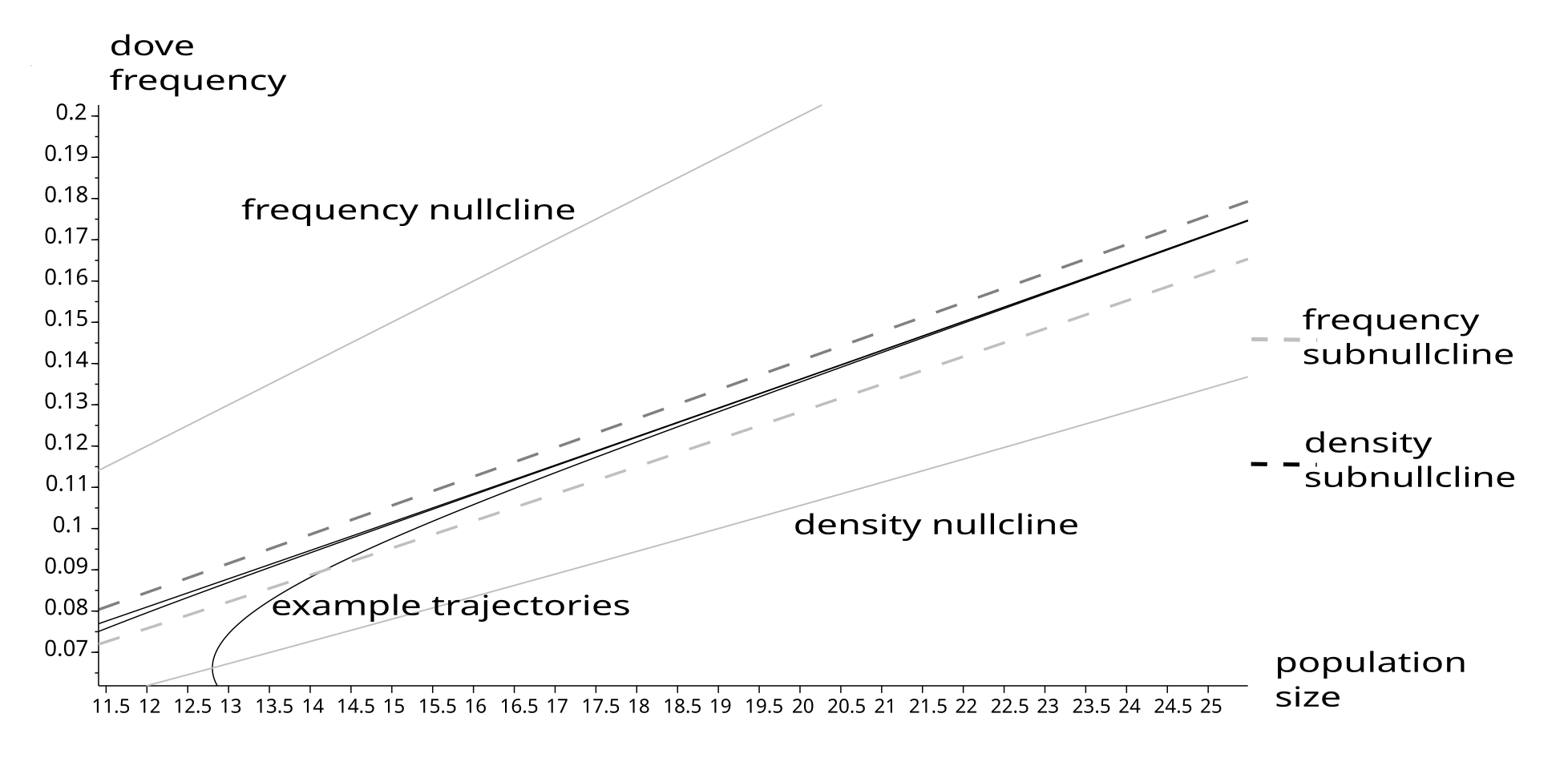}
    \caption{Zoom of the subnullclines. We can see the bundle of
trajectories between subnullclines and another trajectory converging to the
close neighbourhood of the bundle.}
    \label{fig:BundleaOfTrajectories}
\end{subfigure}
\begin{subfigure}[t]{0.49\textwidth}
    \centering
    \includegraphics[width=\linewidth]{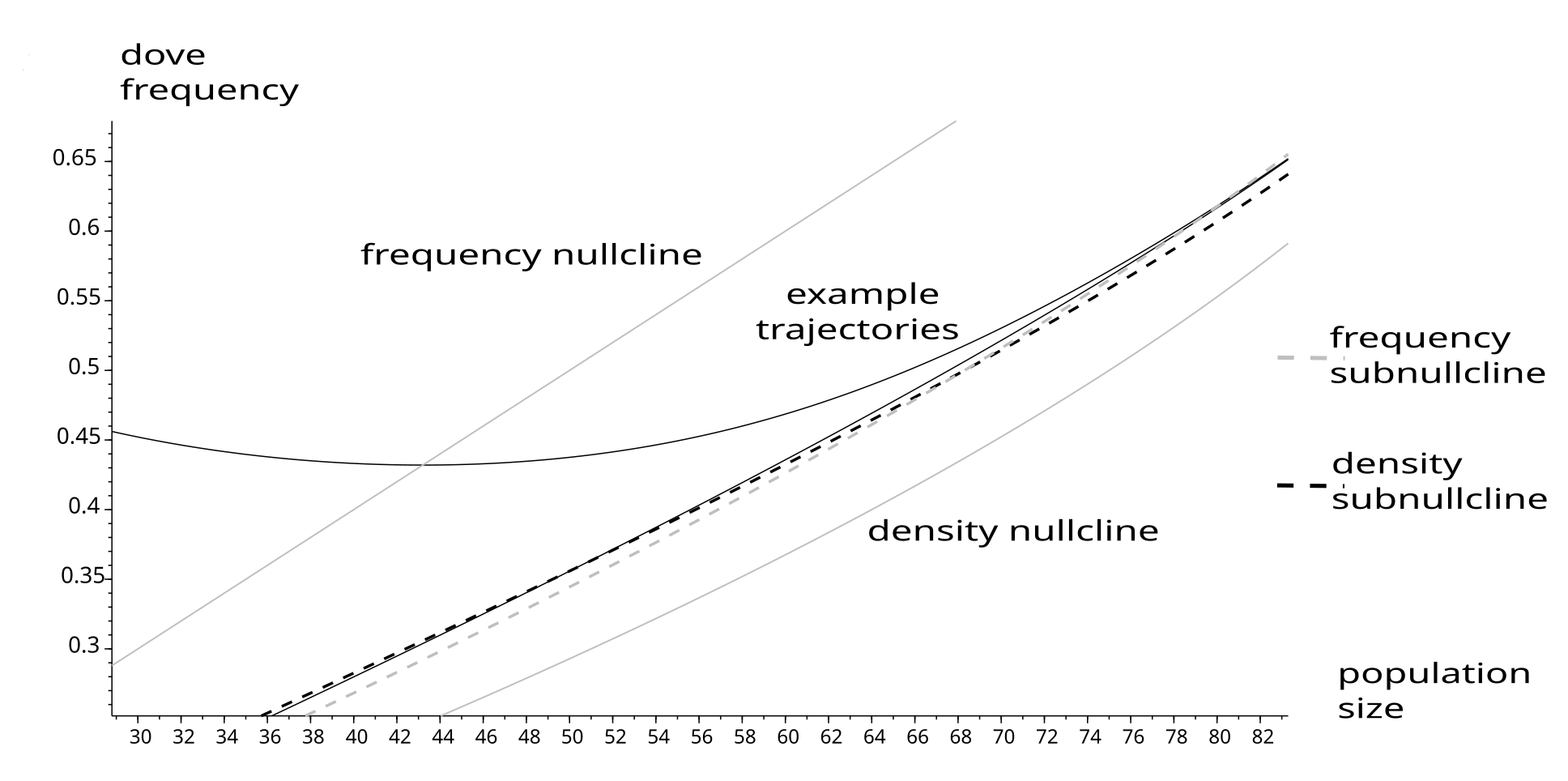}
    \caption{Zoom of the intersection of the subnullclines. The region bounded by subnullclines shrinks fast and the bundle of trajectories falls for a while into the region of attraction.}
    \label{fig:AreaOfIntersection}
\end{subfigure}
\caption{Zoom of the subnullclines from Figure~\ref{fig:main}}
\label{fig:Zoom}
\end{figure}

Figure~\ref{fig:Zoom} shows the zoom on the neighborhood of the subnullclines, in which panel~\ref{fig:BundleaOfTrajectories}
shows the bundle of trajectories running along the subnullclines. Panel~\ref{fig:AreaOfIntersection} however,
shows the area of the intersection of the subnullclines. It shows that the
area limited by subnullclines is not strictly invariant since it can shrink
faster than the trajectories are chasing it. But when the area grows
again, the trajectories enter it again. 

\begin{figure}[htbp!]
\centering
\begin{subfigure}[t]{0.48\textwidth}
    \centering
    \includegraphics[width=\linewidth]{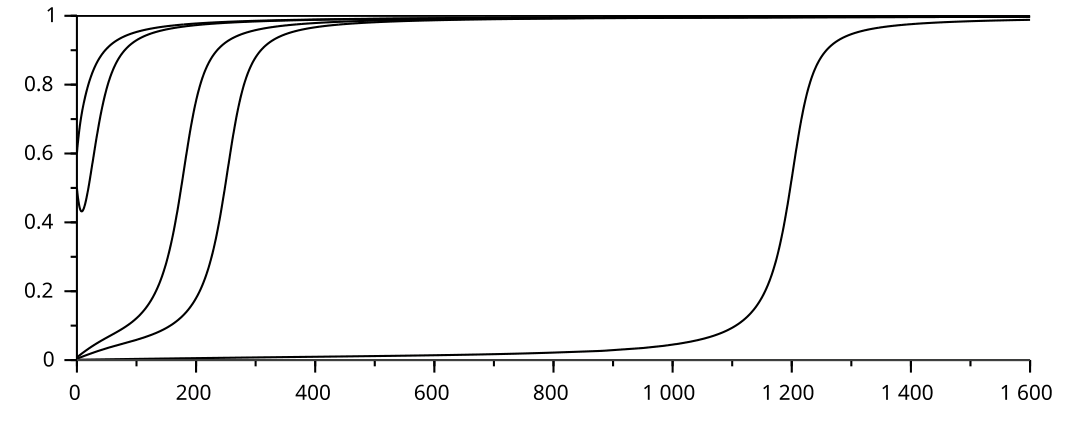}
    \caption{Dove frequency $q_d$}
    \label{fig:3a}
\end{subfigure}
\hfill
\begin{subfigure}[t]{0.48\textwidth}
    \centering
    \includegraphics[width=\linewidth]{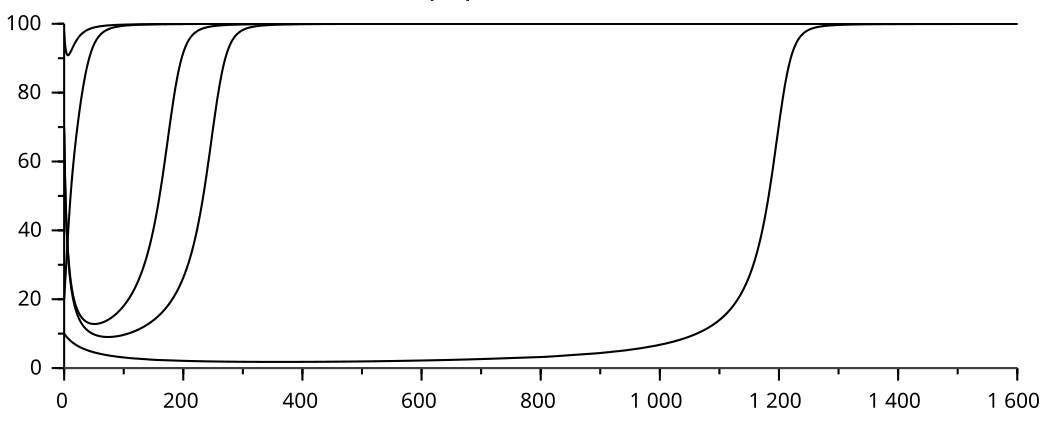}
    \caption{Population size $n$}
    \label{fig:3b}
\end{subfigure}
\caption{Time trajectories show that the convergence rate to the
neighborhood of subnullclines can differ dramatically. After some time, all
trajectories start to behave similarly. This means that they reached the
neighborhood of subnullclines. }
\label{fig:Time Trajectories}
\end{figure}

The time trajectories can be seen in Figure~\ref{fig:Time Trajectories}. Note that the shape of the trajectories contradicts the common belief that \emph{ecological} density dynamics is "fast" while \emph{evolutionary} selection dynamics is "slow" . Both processes arise from demography and are simultaneous. Separation of both processes occurs for greater values of
background vital rates \cite{argbr2}.

\subsection{Subnullclines for detecting the ghost attractor}

When the saddle and node rest points collide and disappear, a phenomenon described as a long transient may occur \cite{Hastings1,Hastings2,Morozov,Morozov2,Koch2024}. One type of long transient occurs when the internal attracting nullclines of the system do not intersect or become disconnected as a result of a bifurcation, but pass close to each other. Between the nullclines, the speed of the flow is extremely slow, since the right-hand sides of both equations are close to zero \cite{Morozov2,Koch2024}. This phenomenon is also referred to as bifurcation memory. The area in the narrow channel that traps the trajectories is called a \emph{ghost attractor}. This is not a specific point or manifold, such as a line, but rather an area that attracts trajectories. In this paper, we present a concept that allows us to locate the ghost attractor in phase space.

\begin{figure}[htbp]
\centering
\begin{subfigure}{0.49\textwidth}
    \centering
    \includegraphics[width=\linewidth]{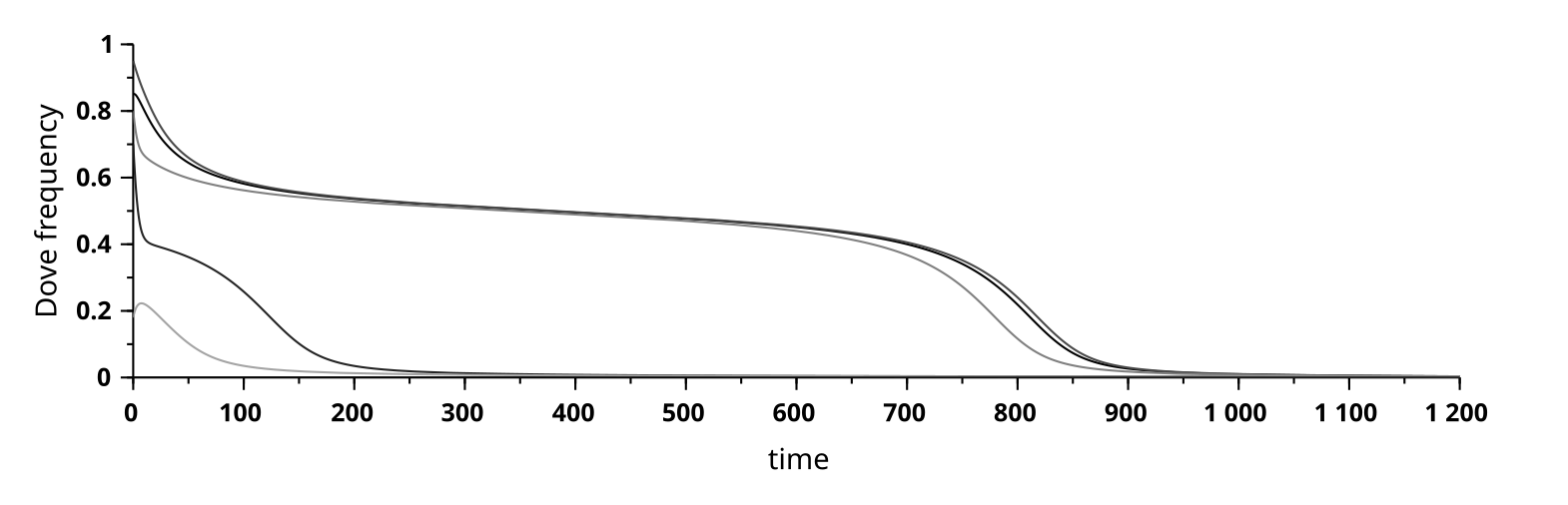}
\end{subfigure}
\begin{subfigure}{0.49\textwidth}
    \centering
    \includegraphics[width=\linewidth]{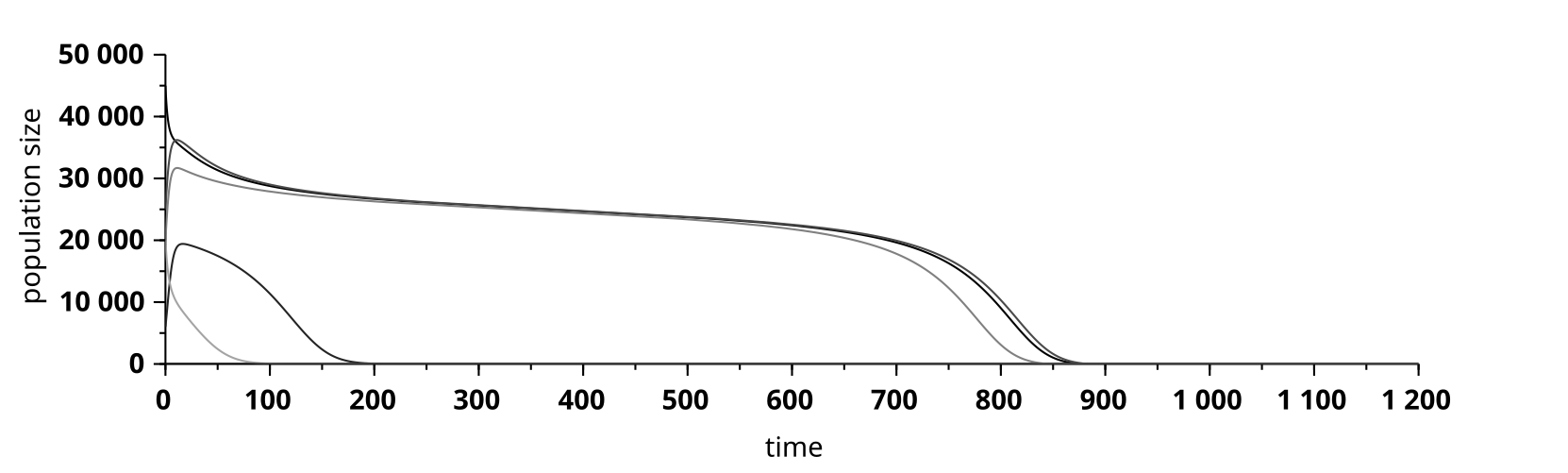}
\end{subfigure}
\caption{Time trajectories for the parameters $F=1$, $d=1$ $\Phi=0$ and $%
\Psi=0.126$. We observe the long transient behaviour. Some trajectories
temporarily slow down, pretending the convergence to the stable equilibrium
followed by the rapid decline.}
\label{LongTrTime}
\end{figure}

\begin{figure}[htbp]
\centering
\begin{subfigure}{0.48\textwidth}
    \centering
    \includegraphics[width=\linewidth]{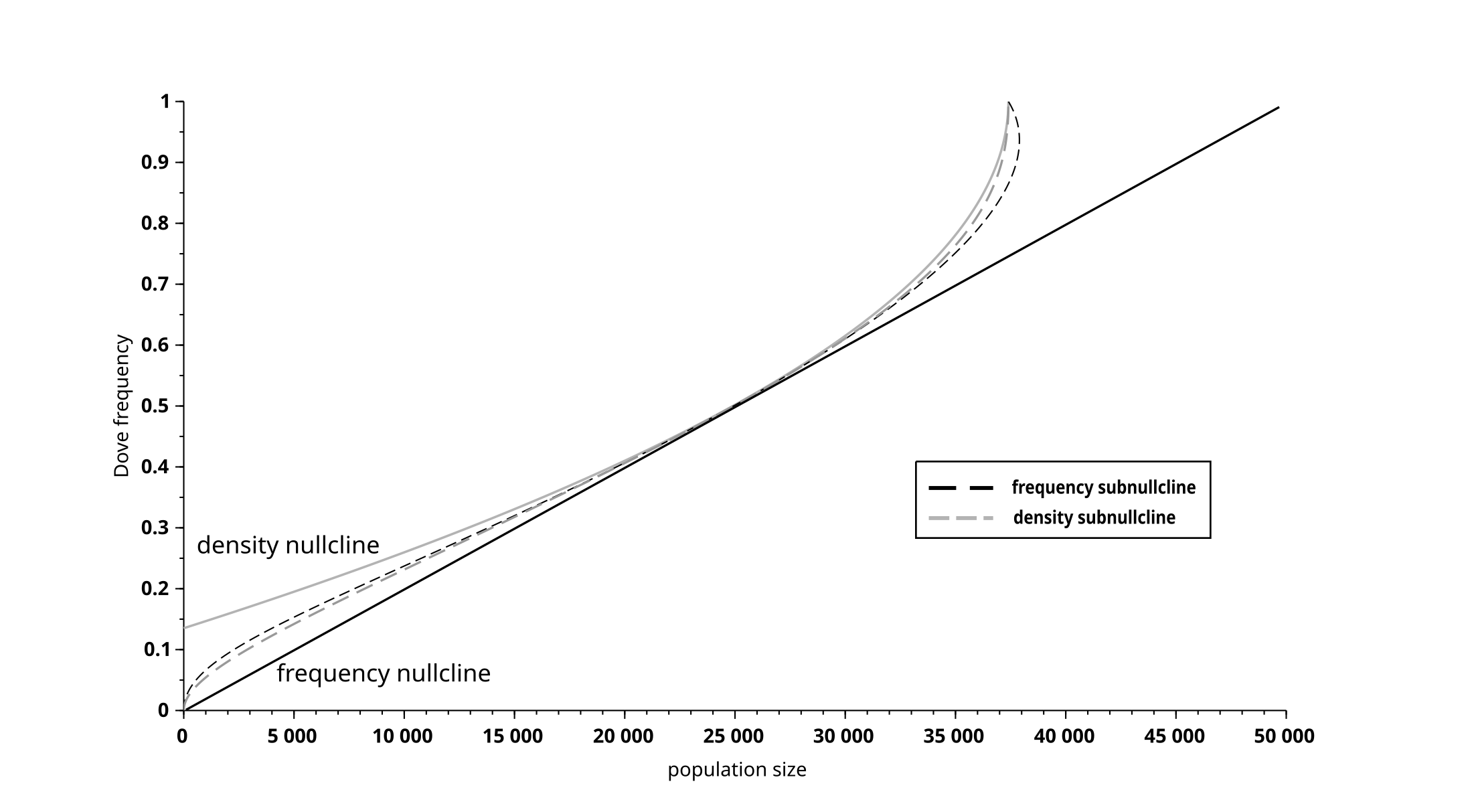}
    \caption{Structure of the nullclines and subnullclines.}
    \label{LongTrPhaseCl}
\end{subfigure}
\hfill
\begin{subfigure}{0.48\textwidth}
    \centering
    \includegraphics[width=\linewidth]{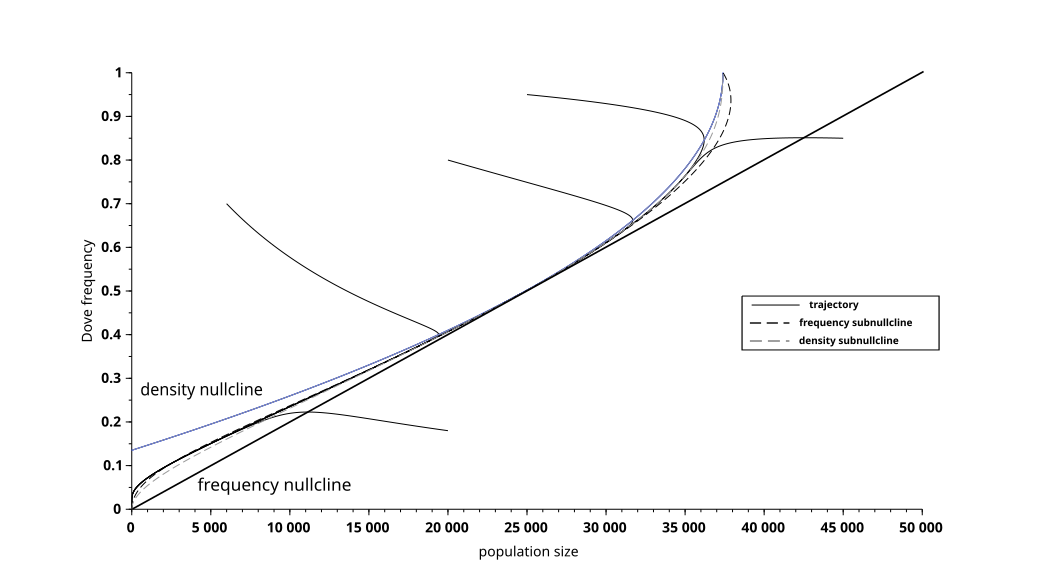}
    \caption{Introduction of sample trajectories.}
    \label{LongTrPhaseTr}
\end{subfigure}
\caption{Plot of the nullclines and the resulting subnullclines for the
example from figure \protect\ref{LongTrTime}.}
\label{fig:}
\end{figure}

\begin{figure}[htbp]
\centering
\begin{subfigure}{0.48\textwidth}
    \centering
    \includegraphics[
    width=\linewidth,
    trim={1.2cm 0.5cm 2cm 2cm},clip]{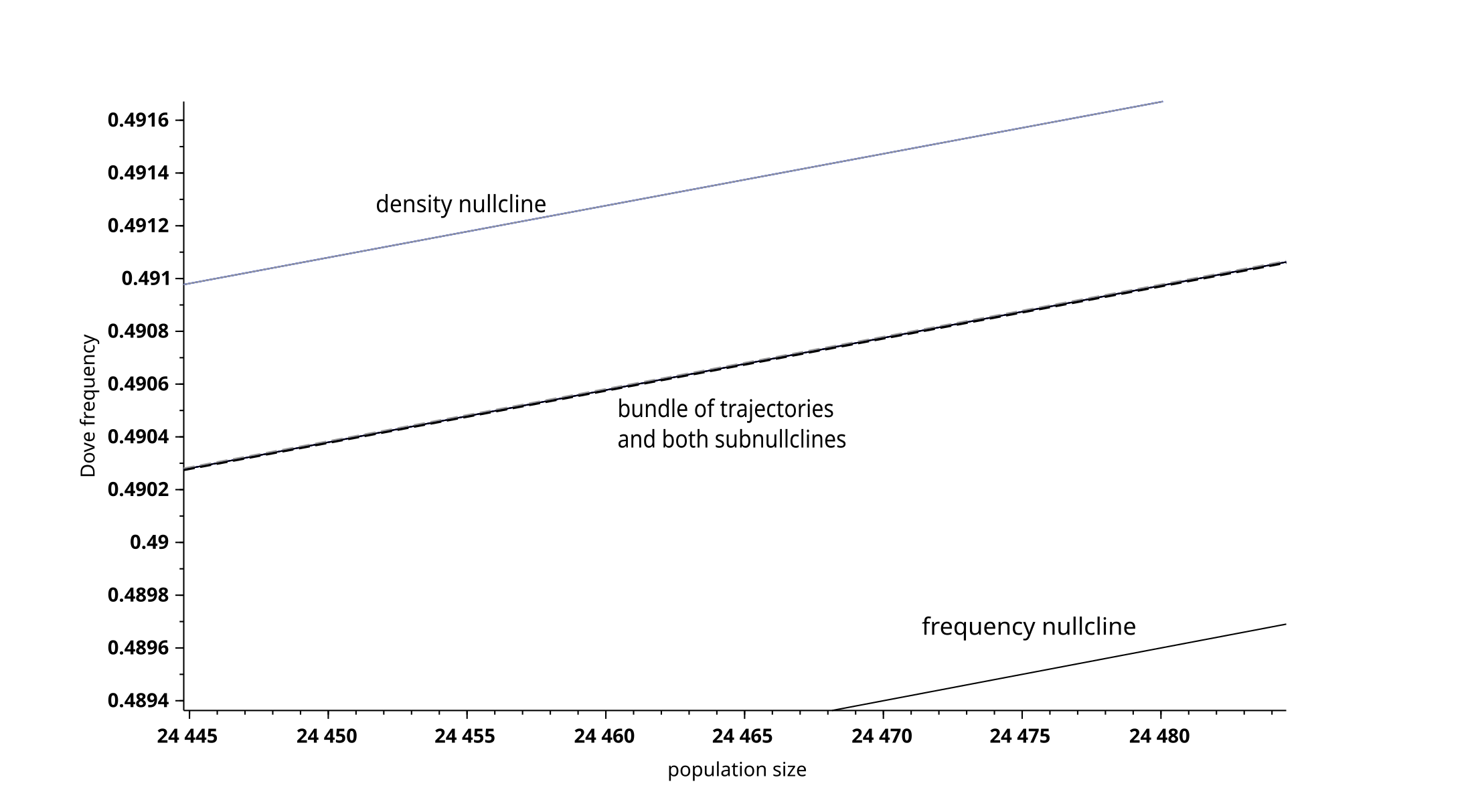}
    \caption{Zoom of the narrow channel between nullclines. The subnullclines in close proximity to the trajectories which form a bundle.}
    \label{LongTrPhaseZ1}
\end{subfigure}
\hfill
\begin{subfigure}{0.48\textwidth}
    \centering
    \includegraphics[
    width=\linewidth,
    trim={1cm 0.6cm 2.5cm 1.5cm},clip]{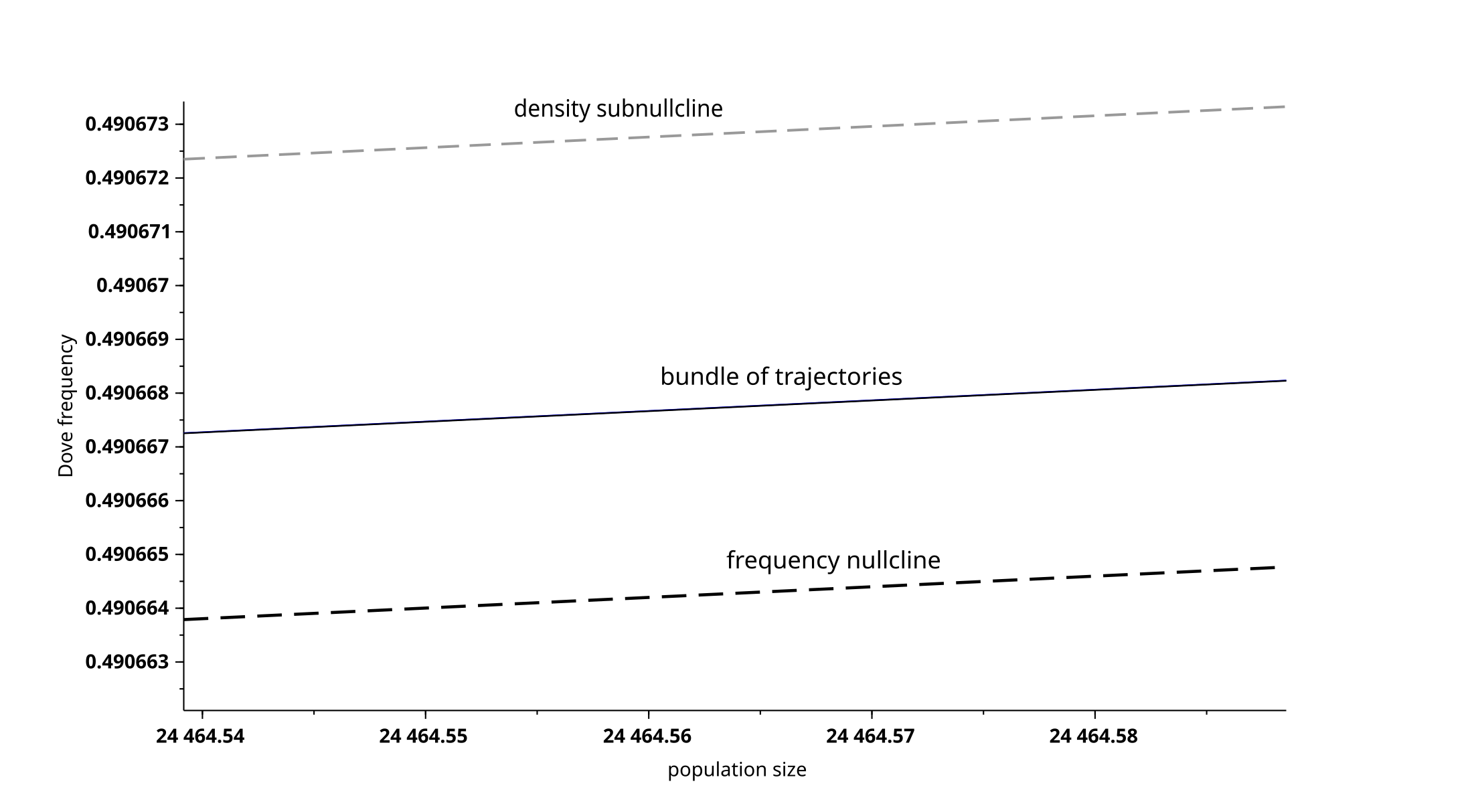}
    \caption{ Bundle of trajectories is significantly narrower
than the size of the channel between subnullclines, which provide extremely accurate approximation in the scale of the whole phase space. }
    \label{LongTrPhaseZ2}
\end{subfigure}
\caption{Zoom of the area between nullclines and subnullclines for figure \protect\ref{LongTrTime}.}
\label{fig:}
\end{figure}

Subnullclines are useful not only in the case of a saddle-node heteroclinic connection, but they also provide an excellent approximation of the ghost attractor. Let us illustrate this with the example shown in Figure~\ref{LongTrTime} wherein some trajectories temporarily slow down, pretending the convergence to the stable equilibrium followed by the rapid decline. Figure~\ref{LongTrPhaseCl} shows the structure of the nullclines and subnullclines, while Figure~\ref{LongTrPhaseTr} also includes example trajectories. Figure~\ref{LongTrPhaseZ1} shows a zoomed view of the narrow channel between the nullclines, which leads to the ghost attractor. We can see that the bundle of trajectories and both subnullclines are extremely close to each other, even compared with the scale of the channel. Therefore, the subnullclines provide a very accurate approximation of the ghost attractor. However, Figure~\ref{LongTrPhaseZ2} shows that the bundle of trajectories is even narrower compared with the scale of the distance between the subnullclines. From the perspective of the general phase portrait, however, the subnullclines can still be regarded as an excellent approximation. The differences are observable only at microscopic scales. When the channel between the nullclines becomes wider, the duration of the long transients decreases, as can be seen in the supplementary examples from Appendix~\ref{Appendix 5}. 

\begin{figure}[htbp]
\centering
\begin{subfigure}{0.49\textwidth}
    \centering
    \includegraphics[width=10cm]{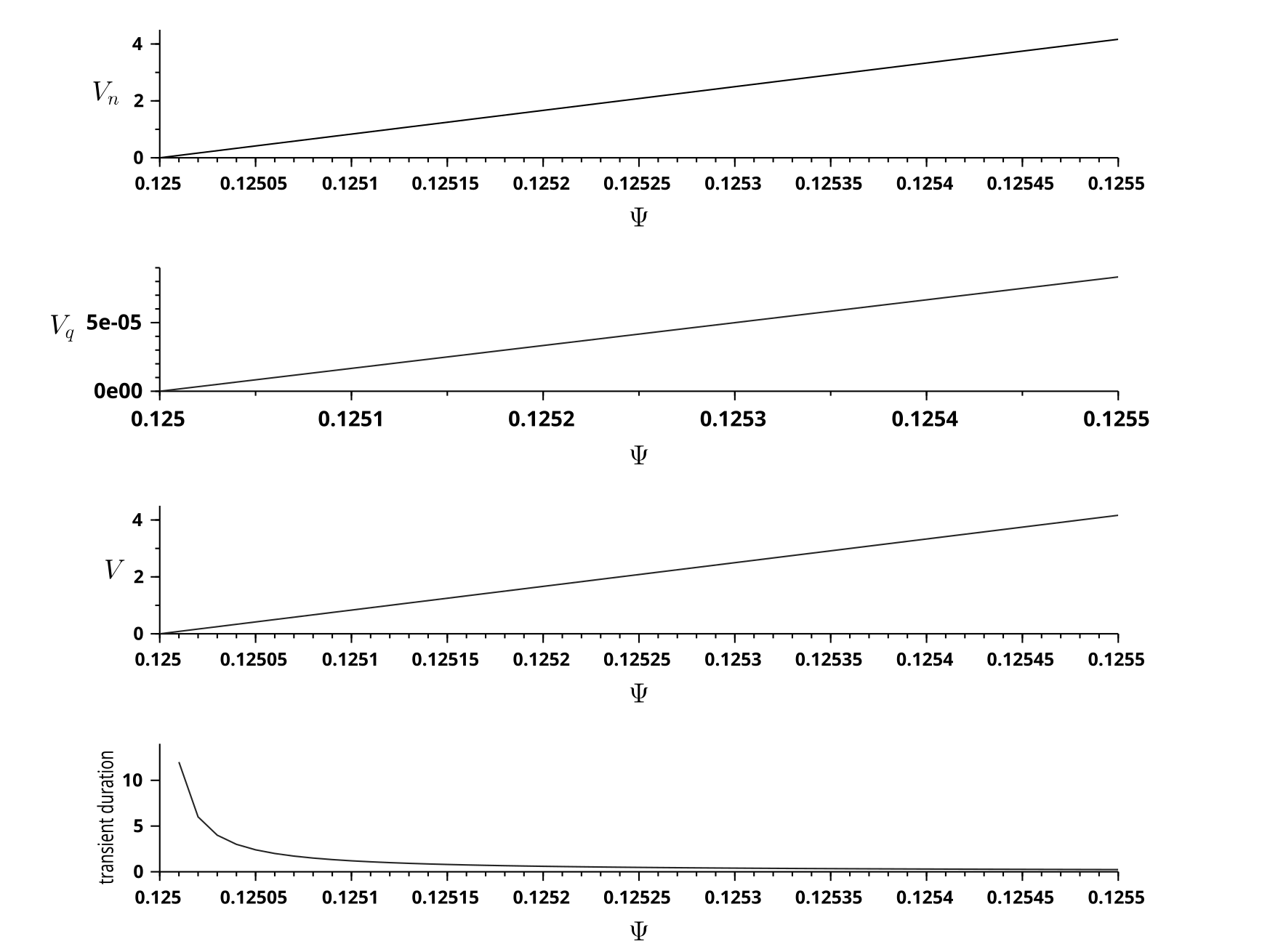}  
    \caption{Plot of the vertical, horizontal, and general velocities as functions of the parameter $\Psi$, evaluated at the point of intersection of both subnullclines. The respective velocities $V_q$, $V_h$ and $V$ decrease nearly linearly and the transient duration $1/V$ is proportional to an inverse power law.}
    \label{SPEED}
\end{subfigure}
\hfill
\begin{subfigure}{0.49\textwidth}
    \centering
    \includegraphics[width=\linewidth]{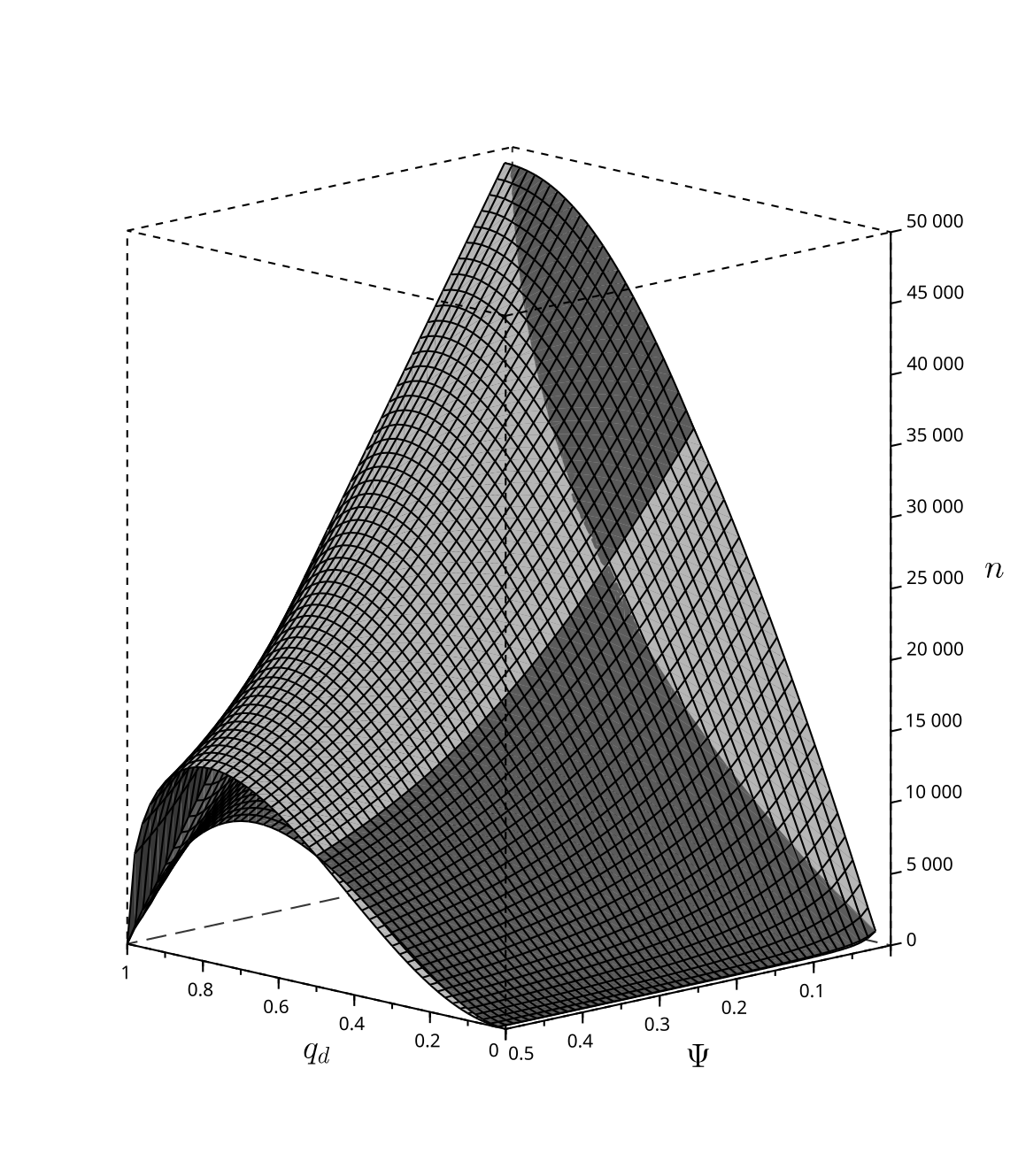}
    \caption{Plot of the subnulllcines as the function of the parameter $\Psi$. The bifurcation is visible in the structure of the intersections of the
subnullclines }
\label{3Dsub}
\end{subfigure}
\caption{Plot of the velocities and subnullclines as a function of the parameter $\Psi$.}
\label{fig:}
\end{figure}

We can use the point $(\breve{q}_{d},\mathcal{S}^{\varepsilon}_q(\breve{q}_{d}))$, common to both subnullclines \eqref{SubInter}, to evaluate how the rate of the flow, described by the right-hand sides of equations \eqref{repq} and \eqref{repn}, changes as a function of the parameter ($\Psi$). This change is associated with the changing width of the channel between the nullclines. We numerically evaluate the quantities
\[
V_{q}=\left\vert g(\breve{q}_{d},\mathcal{S}^{\varepsilon}_n(\breve{q}_{d}))\right\vert,
\qquad
V_{n}=\left\vert f(\breve{q}_{d},\mathcal{S}^{\varepsilon}_n(\breve{q}_{d}))\right\vert,
\qquad
V=
\sqrt{
\left[ g(\breve{q}_{d},\mathcal{S}^{\varepsilon}_n(\breve{q}_{d}))\right]^{2}
+
\left[ f(\breve{q}_{d},\mathcal{S}^{\varepsilon}_n(\breve{q}_{d}))\right]^{2}
},
\]
which describe, respectively, the vertical, horizontal and total rates of movement at the intersection point of the subnullclines. Figure~\ref{SPEED} shows these quantities as functions of the parameter ($\Psi$). Although the explicit forms of ($V_q$), ($V_n$) and ($V$) are too complicated to be shown in the paper, their numerical behaviour supports the claim in \cite{Morozov2} that the duration of the long transient is characterized by an inverse power law.

\begin{figure}[tbp]
\centering

\end{figure}

\FloatBarrier

\subsection{Subnullclines to determine the resilience and resistance thresholds}

In this subsection, we focus on the impact of the strategically neutral seasonal periodic mortality factors on the dynamics of strategy frequencies.
Subnullclines act as approximate transient attractors, since they constitute the border between repellence from and attraction to the focal nullcline. This is important from the point of view of system resilience, \cite{ResHolling1,ResHolling2,ResReed,ResKrak,ResGund,ResMeyer}, defined as the propensity to absorb a perturbation that alters the state while maintaining the functions of the system. Probably every population on Earth is affected by some periodic factors. In the so-called ``flow-kick'' approach \cite{ResMeyer}, perturbation is represented by a discrete shift of the population state followed by a continuous return phase, modeled by a standard autonomous differential equation. In contrast, we incorporate the perturbation as an explicit part of the model.
 


Different periodic mortality factors are common in nature. We can modify our
Hawk--Dove example by adding an external periodic mortality factor,
$\alpha \left(1+\sin \left(\frac{2\pi}{\theta}t\right)\right)$, where
$\alpha$ describes the mortality amplitude and $\theta$ is the duration of
the season. Then, the resulting system is: 
\begin{equation}
\frac{dq_{d}}{dt}
=
\frac{1}{2}q_{d}(t)\bigl(1-q_{d}(t)\bigr)
\left[
\bigl(1-q_{d}(t)\bigr)d
-
\left(1-\frac{n(t)}{K}\right)F
\right].
\label{freqprey}
\end{equation}

\begin{equation}
\frac{dn}{dt}
=
n(t)
\left[
\left(\frac{1}{2}F+\Phi\right)
\left(1-\frac{n(t)}{K}\right)
-
\left\{
\frac{1}{2}d\bigl(1-q_{d}(t)\bigr)^{2}
+\Psi
+\alpha\left(1+\sin\left(\frac{2\pi}{\theta}t\right)\right)
\right\}
\right].
\label{prey}
\end{equation}

Then, the perturbation acting as additional periodic mortality factor leads to the movement of the density nullcline. This induces oscillations in the system, since the trajectories follow the moving attracting surface. Therefore the subnullclines should provide approximate limiting borders for those oscillations.

\subsubsection{ Case with additional seasonal adult mortality leading to the
game with oscillating rewards}

\begin{figure}[htbp!]
\centering
\begin{subfigure}{0.55\textwidth}
    \centering
    \includegraphics[width=\linewidth]{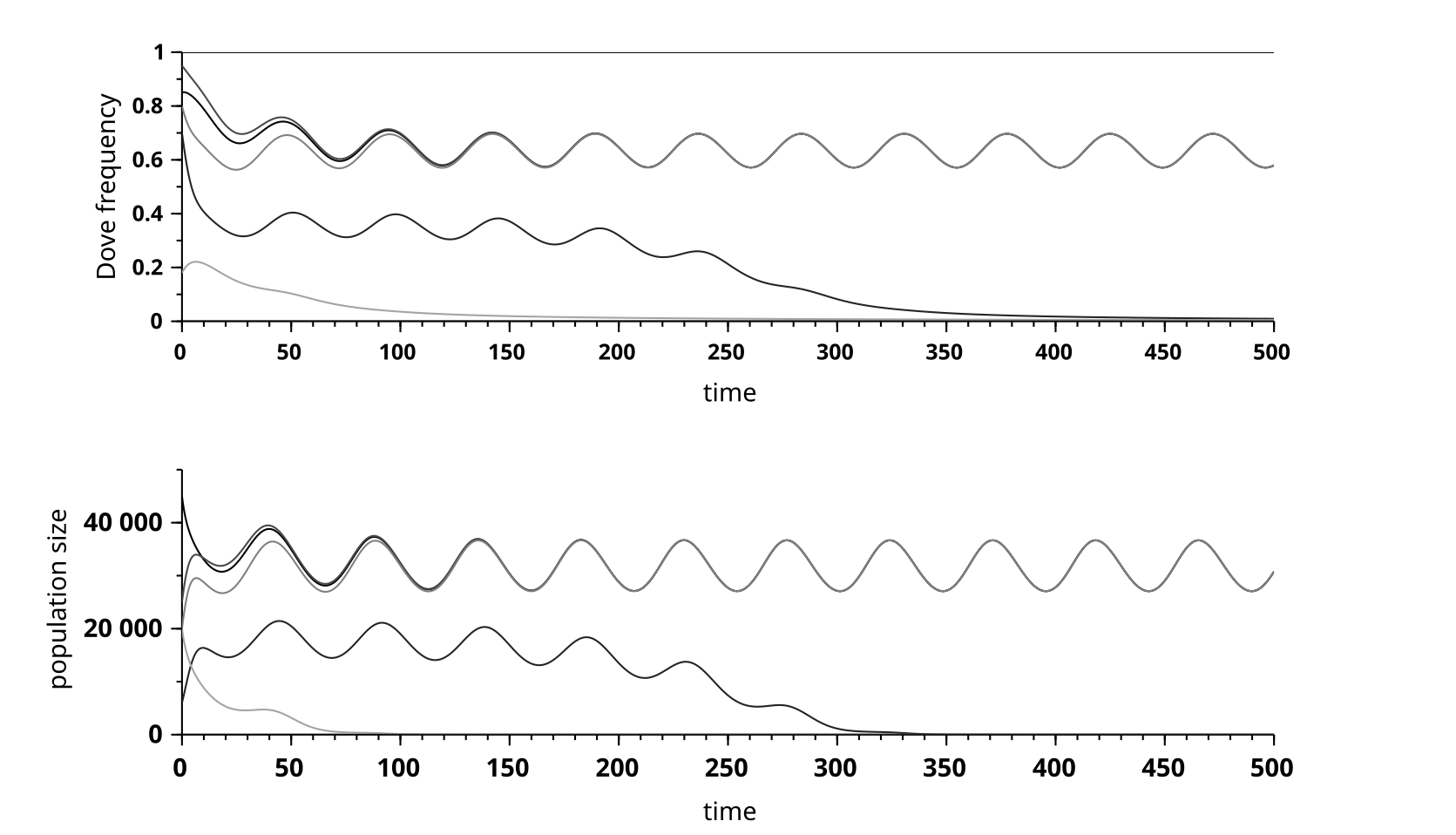}
    \caption{Temporal dynamics for frequency and density equations.}
\label{ResTime}
\end{subfigure}
\hfill
\begin{subfigure}{0.4\textwidth}
    \centering
    \includegraphics[width=\linewidth]{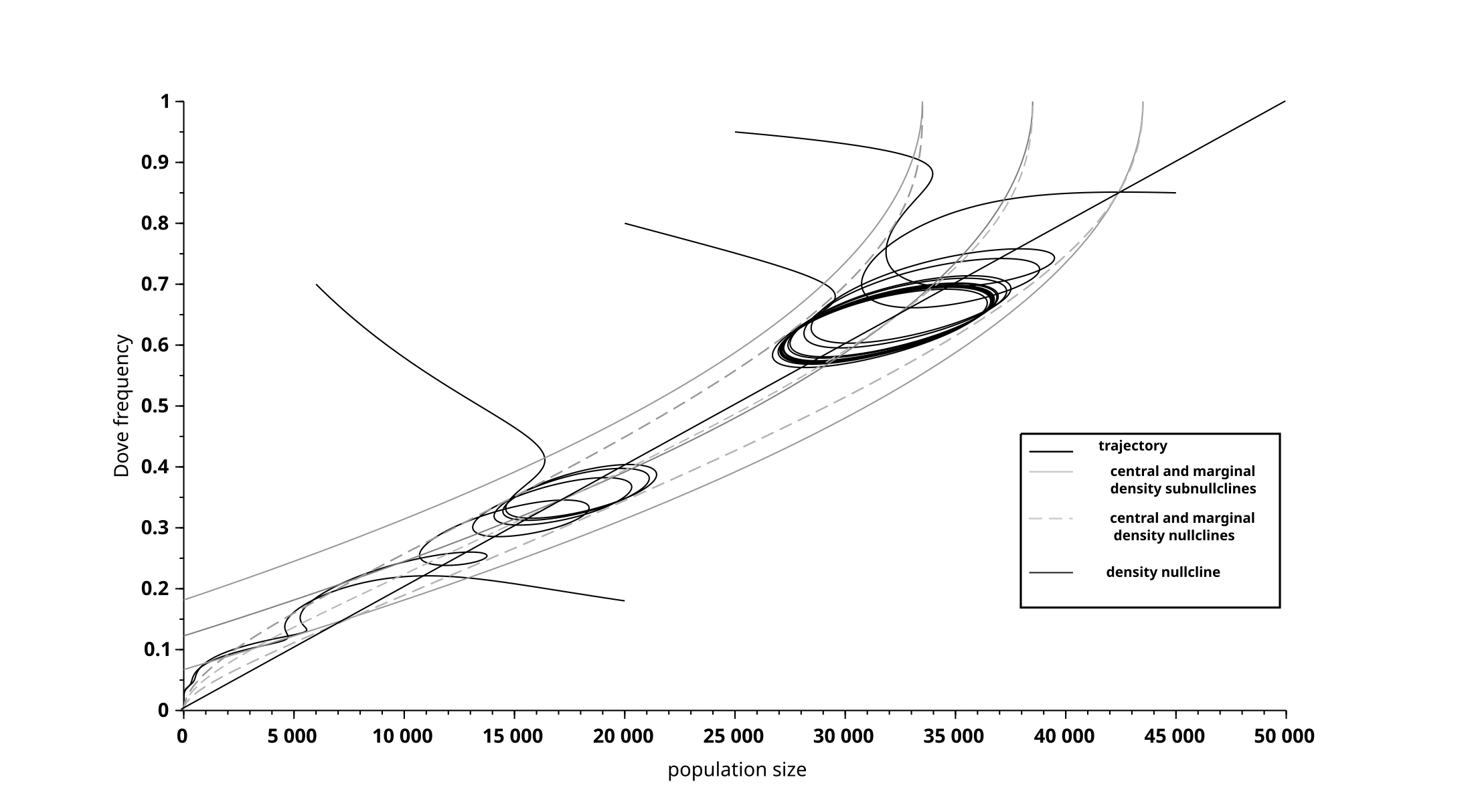}
\caption{Phase portrait corresponding to Figure~\protect\ref{ResTime}. The rest points of the unperturbed system are transformed into stable and unstable cycles. The subnullclines bound the range of oscillations and can therefore be interpreted as resilience or resistance thresholds. They mark the boundary between the absorption of a perturbation and a strong reaction that prevents its further propagation.}
\label{ResPhase}
\end{subfigure}
\caption{Time trajectories for the parameters $F=1$, $d=1$, $\Phi=0$, and
$\Psi=0.065$, under the pressure of a periodic factor with amplitude $0.05$
and period $2\pi\theta=7.5$.}
\label{fig:FigureRes}
\end{figure}

Let us examine the sensitivity of our system to this type of perturbation. Numerical simulations reveal interesting response patterns under periodic mortality. Figure~\ref{fig:FigureRes} shows that, in the system with temporarily vanishing rest points caused by cyclic disconnection of the nullclines, the system maintains its quantitative properties. The original rest points are transformed into stable and unstable cycles. We observe that the subnullclines of the marginal density nullclines approximately limit the range of oscillations induced by the periodic factor. Therefore, they can be interpreted as thresholds between resilience, understood as the absorption of the perturbation by the system and resistance, understood as the opposition of the system to the propagation of the perturbation.

\begin{figure}[htbp!]
\centering
\begin{subfigure}{0.8\linewidth}
    \centering
    \includegraphics[
    width=\linewidth]{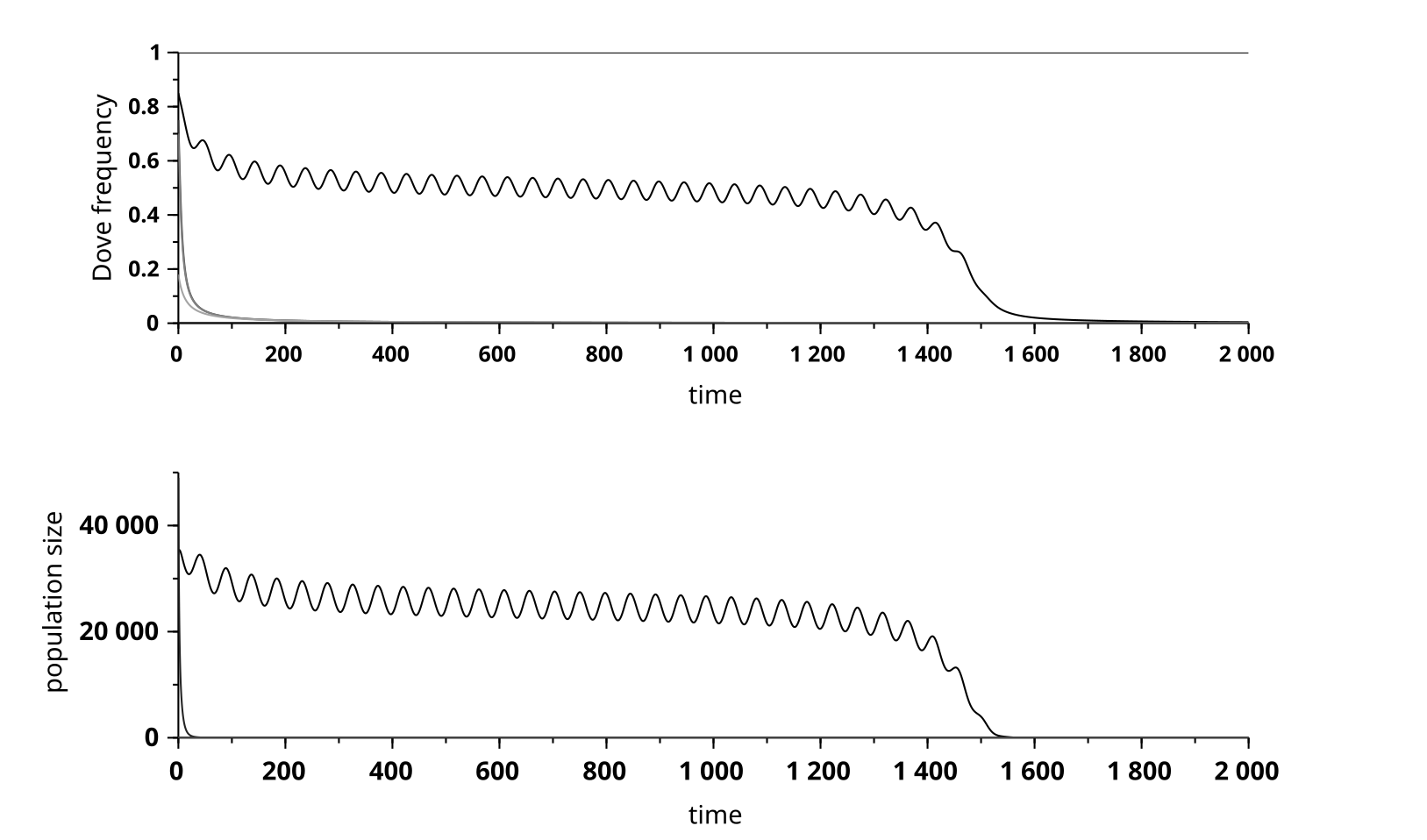}
\caption{Temporal dynamics for frequency and density equations.}
\label{LTPerTime}
\end{subfigure}
\begin{subfigure}{0.8\linewidth}
    \centering
    \includegraphics[width=\linewidth]{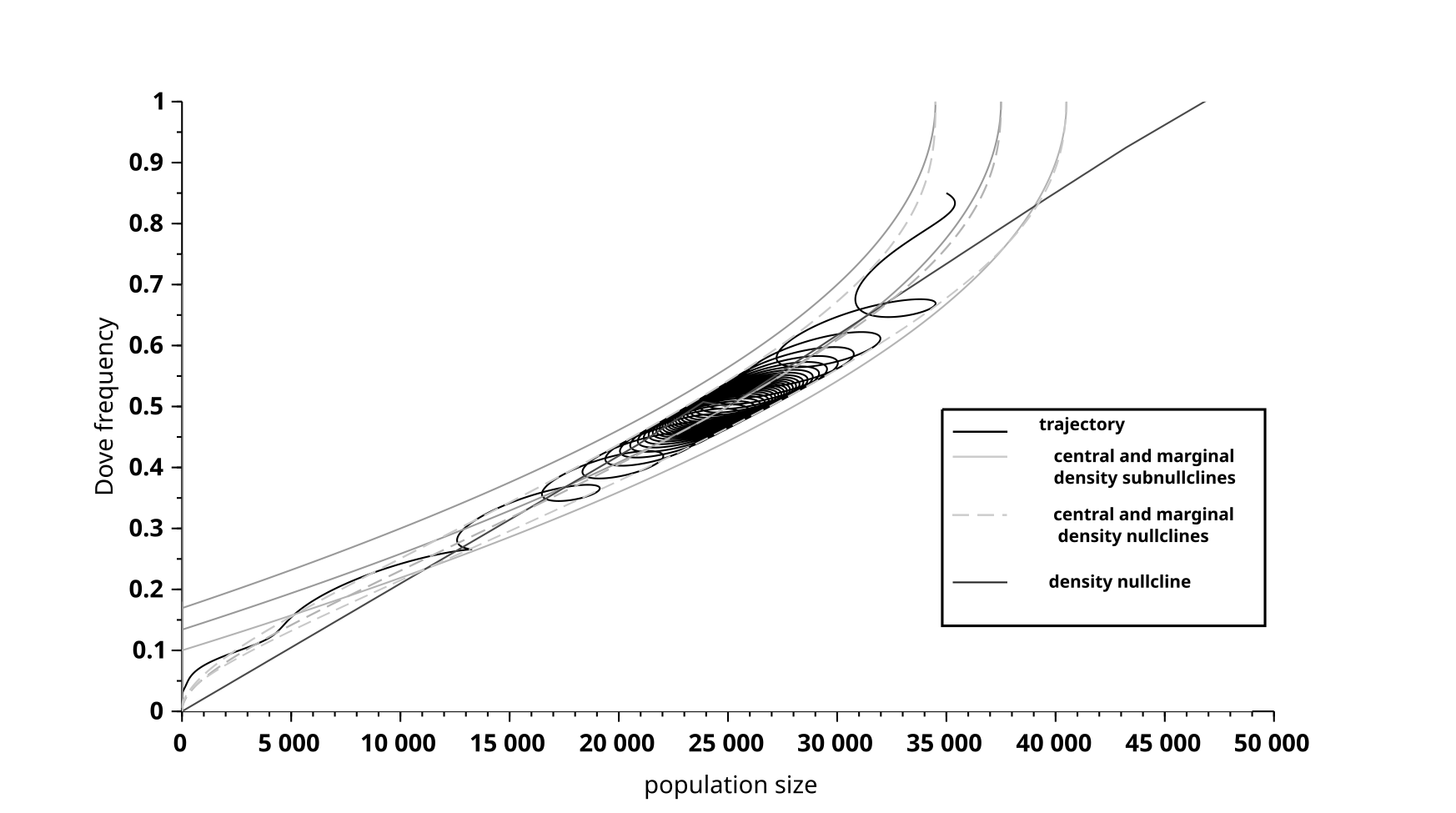}
\caption{Phase portrait corresponding to figure \ref{LTPerTime}. Oscillations of the perturbed ghost attractor are limited by
subnullclines.}
\label{LTPerPhase}
\end{subfigure}
\caption{Time trajectories for the parameters  $F=1$, $d=1$ $\Phi=0$ and $%
\Psi=0.095$, under the pressure of a periodic factor with amplitude $0.03$
and period $2\pi\theta=7.5$.}
\label{fig:FigureLT}
\end{figure}

A similar situation occurs in the case of the perturbed ghost attractor, presented in Figure~\ref{fig:FigureLT}. The marginal subnullclines form a channel that bounds the oscillating trajectories. Although this barrier is approximate, it provides a good description of the observed dynamics.


\section{Discussion}

The eco-evolutionary dynamics converge to the interior of the area bounded by the intersecting frequency and density nullclines. They are attracted by a structure consisting of two intersecting frequency and density subnullclines, defined as manifolds where the slope of the flow is equal to the slope of the respective frequency or density nullcline. Trajectories converge to a close neighbourhood of the heteroclinic orbit connecting the stable and unstable intersections of the nullclines. This orbit is attracted to the neighbourhood of the subnullclines. Then, the trajectories converge to the stable intersection along the subnullclines. In the case of an attracting frequency nullcline and two intersections, the stability of the intersection can be inferred from the sign of the right-hand side of the frequency equation on the stable density nullcline. This is a simple outcome of Theorem 2 from \cite{argbr3}, but it simplifies the calculations.

In addition, it is shown that the eco-evolutionary dynamics discussed above are highly sensitive to external factors, such as seasonal mortality, which can lead to cycling behaviour around the stable state, alter the convergence patterns and also generate complex cycling behaviour. This is important because real populations are not isolated systems, but parts of larger and more complex ecosystems. Our results show that the impact of other elements of the environment or ecosystem can substantially alter the behaviour of a particular population. This perspective emphasizes the impact and role of ecological factors in the processes of natural selection. Note that, in this case, common simplifying approaches, such as the assumption of weak selection, become problematic/complicated and should therefore be carefully reexamined.


\section*{Acknowledgments}
 This work is generously supported by grant OPUS 2020/39/B/NZ8/03485 awarded to Krzysztof Argasinski by Polish National Science Center and the studentship awarded to Manjyot Singh Bedi by City St. George's, University of London, for which we are very grateful.

\appendix

\section{Appendix 1}\label{Appendix 1}

Recall that Hawk's fertility payoff is 
\begin{equation*}
V_{h}=q_{h}0.5F+(1-q_{h})F=\left( 1-0.5q_{h}\right) F=0.5 F\left(1+q_d\right)
\end{equation*}
\ and Dove's fertility payoff is 
\begin{equation*}
V_{d}=0.5Fq_d+0q_{h}=(0.5F)q_{d},
\end{equation*}
Then, the average fertility payoff is 
\begin{eqnarray*}
\bar{V}
&=&q_{h}V_{h}+(1-q_{h})V_{d}=q_{h}^{2}0.5F+q_{h}(1-q_{h})F+(1-q_{h})^{2}0.5F
\\
&=&\left[ q_{h}^{2}0.5+q_{h}(1-q_{h})+(1-q_{h})^{2}0.5\right] F \\
&=&\left[ \left( q_{h}^{2}+(1-q_{h})^{2}\right) 0.5+q_{h}(1-q_{h})\right] F
\\
&=&\left[ \left( q_{h}^{2}+1-2q_{h}+q_{h}{}^{2}\right) 0.5+q_{h}(1-q_{h}) %
\right] F \\
&=&\left[ \left( 1-2q_{h}+2q_{h}{}^{2}\right) 0.5+q_{h}(1-q_{h})\right] F \\
&=&\left[ 0.5-q_{h}+q_{h}{}^{2}+q_{h}-q_{h}^{2})\right] F=0.5F,
\end{eqnarray*}

which means that every game round has a winner. Mortality payoffs are $%
D_{h}=(0.5d)q_{h}$ and $\ D_{d}=0$. The average mortality is $\bar{D}
=q_{h}D_{h}+q_{d}D_{d}=0.5d(1-q_{d})^{2}$. Then, we derive the
replicator equations for Doves, which is simpler due to the lack of
Dove vs Dove mortality. Assume that $q_{d}=1-q_{h}$, then
\begin{eqnarray*}
\frac{dq_{d}}{dt} &=q_{d}(t)&\left[ \left( 1-\frac{%
n(t) }{K}\right) \left(0.5Fq_{d}-0.5F\right)-\left( 0-0.5d\left( 1-q_{d}(t)\right) ^{2}\right)
\right] \\
\frac{dn}{dt} &=n(t)&\left[ \left( 1-\frac{n(t)}{K}
\right)\left( 0.5F+\Phi \right)  -\left( 0.5d\left(1-q_{d}(t)\right) ^{2}+\Psi \right) \right] .
\end{eqnarray*}

After rearranging we have:
\begin{eqnarray}
\frac{dq_{d}}{dt} &=&0.5q_{d}(t)\Big( 1-q_{d}(t)\Big) \left[ \Big(
1-q_{d}(t)\Big) d-\left( 1-\frac{n(t)}{K}\right) F\right] \\
\frac{dn}{dt} &=&n(t)\left[ \Big( 0.5F+\Phi \Big) \left( 1-\frac{n(t)}{K}
\right) -\Big( \big( 1-q_{d}(t)\big) ^{2}0.5d+\Psi \Big) \right] .
\end{eqnarray}

\section{Appendix 2}\label{Appendix 2}

Below, we calculate the nullclines and the rest points for the classic Hawk-Dove
game in our eco-evolutionary setup. Nullcline for the dove frequency will be,
\begin{gather}
\Big( 1-q_{d}(t)\Big) d-\left( 1-\frac{n(t)}{K}\right) F=0 \\
q_{d}(t)=1-\left( 1-\frac{n(t)}{K}\right) \frac{F}{d}.
\end{gather}

Nullcline for the population size will be,
\begin{gather}
\left( 0.5F+\Phi \right) \left( 1-\frac{n(t)}{K}\right) -\left[ \left(
1-q_{d}(t)\right) ^{2}0.5d+\Psi \right] =0 \\[0.3em]
1-\frac{n(t)}{K}=\frac{\left( 1-q_{d}(t)\right) ^{2}0.5d+\Psi }{0.5F+\Phi }
\label{logisticterm} \\[0.3em]
n(t)=\left[ 1-\frac{\left( 1-q_{d}(t)\right) ^{2}0.5d+\Psi }{0.5F+\Phi } %
\right] K.
\end{gather}

The rest points will be intersections of both nullclines. After substitution
of (\ref{logisticterm}) into (\ref{qnullcline}) we obtain,
\begin{equation*}
q_{d}(t)=1-\Big[\frac{\left( 1-q_{d}(t)\right) ^{2}0.5d+\Psi }{0.5F+\Phi }\Big]\frac{F 
}{d}.
\end{equation*}

Note that for $\Psi =\Phi =0$ the above equation reduces to 
\begin{gather*}
q_{d}(t)=1-\left( 1-q_{d}(t)\right) ^{2} \\
\left( 1-q_{d}(t)\right) -\left( 1-q_{d}(t)\right) ^{2}=0 \\
q_{d}(t)\left( 1-q_{d}(t)\right) =0.
\end{gather*}

Therefore, without the background vital rates, we have only two rest points
for pure strategies. Note that this is independent from the values of $F$
and $d$. In general, we have an equation 
\begin{eqnarray*}
q_{d}(t) &=&1- \left[ \frac{\left( 1-q_{d}(t)\right) ^{2}0.5d+\Psi }{\left(
0.5F+\Phi \right) } \right ]\frac{F}{d} \\
\left( 1-q_{d}(t)\right) \left( 0.5F+\Phi \right) &=&\left[ \left(
1-q_{d}(t)\right) ^{2}0.5d+\Psi \right] \frac{F}{d} \\
\left( 0.5F+\Phi \right) -q_{d}(t)\left( 0.5F+\Phi \right) &=&\left(
1-q_{d}(t)\right) ^{2}0.5F+\frac{F\Psi }{d} \\
\left( 0.5F+\Phi \right) -q_{d}(t)\left( 0.5F+\Phi \right) &=&\left(
1-2q_{d}(t)+q_{d}(t)^{2}\right) 0.5F+\frac{F\Psi }{d} \\
\left( 0.5F+\Phi \right) -q_{d}(t)\left( 0.5F+\Phi \right)
&=&0.5F-q_{d}(t)F+q_{d}(t)^{2}0.5F+\frac{F\Psi }{d} \\
0.5Fq_{d}(t)^{2}+\left( \Phi -0.5F\right) q_{d}(t)+\frac{F\Psi }{d}-\Phi
&=&0.
\end{eqnarray*}

Then $\Delta =\left( \Phi -0.5F\right) ^{2}-2F\left( \frac{F\Psi }{d}-\Phi
\right) $ and the frequencies of the restpoints are

\begin{eqnarray*}
\check{q}_{d} &=&0.5-\frac{\Phi -\sqrt{\left( \Phi -0.5F\right)
^{2}-2F\left( \frac{F\Psi }{d}-\Phi \right) }}{F} \\
\hat{q}_{d} &=&0.5-\frac{\Phi +\sqrt{\left( \Phi -0.5F\right) ^{2}-2F\left( 
\frac{F\Psi }{d}-\Phi \right) }}{F}
\end{eqnarray*}

The stability of the rest point can now be inferred from the sign of the frequency component ($g(q,n)$) from \eqref{Eq:FrequencyRep} along the density nullcline. Substituting the density nullcline into the right-hand side of the frequency equation gives:
\begin{equation}
0.5q_{d}(t)\left( 1-q_{d}(t)\right) \left( \left( 1-q_{d}(t)\right) d-\dfrac{
\left( 1-q_{d}(t)\right) ^{2}0.5d+\Psi }{0.5F+\Phi }F\right)  \label{stab}
\end{equation}
The last bracketed term can be presented as 
\begin{eqnarray*}
&&\left(\left( 1-q_{d}(t)\right) d-\dfrac{\left( 1-q_{d}(t)\right) ^{2}0.5d+\Psi }{
0.5F+\Phi }F \right) \\
&\equiv&\left( 1-q_{d}(t)\right) d\left[ 1-\left( 1-q_{d}(t)\right) \dfrac{0.5F}{
0.5F+\Phi }\right] +\dfrac{\Psi }{0.5F+\Phi }F
\end{eqnarray*}

Since $\left( 1-q_{d}(t)\right) \leq 1$ and $\dfrac{0.5F}{0.5F+\Phi }<1$ we
have that 
\begin{equation*}
1-\left( 1-q_{d}(t)\right) \dfrac{0.5F}{0.5F+\Phi }\geq 0
\end{equation*}

Since \eqref{stab} is always positive, it means that the
rest point $\hat{q}_{d}$ is stable.

\section{Appendix 3}\label{Appendix 3}

Since 
\begin{equation*}
\left( q_{d}(t)-q_{d}^{2}(t)\right) \left( 1-q_{d}(t)\right) =\left(
q_{d}(t)-q_{d}^{2}(t)-q_{d}^{2}(t)+q_{d}^{3}(t)\right) ,
\end{equation*}

right hand sides of the replicator equations (\ref{repq},\ref{repn}) can be
presented in the form 
\begin{eqnarray*}
\frac{dq_{d}}{dt} &=&g(n,q)=0.5\left( \left(
q_{d}(t)-2q_{d}^{2}(t)-q_{d}^{3}(t)\right) d-\left(
q_{d}(t)-q_{d}^{2}(t)\right) \left( 1-\frac{n(t)}{K}\right) F\right) \\
\frac{dn}{dt} &=&f(n,q)=\left( 0.5F+\Phi \right) \left( n(t)-\frac{n(t)^{2}}{
K}\right) -n(t)\left[ \left( 1-2q_{d}(t)+q_{d}(t)^{2}\right) 0.5d+\Psi %
\right] .
\end{eqnarray*}

Derivative of the frequency nullcline (\ref{qnullcline}) equals $\dfrac{F}{%
dK }$. Then, on the subnullcline will be satisfied equation for equality of
slopes

\begin{equation}
\frac{g(x,y)}{f(x,y)}=\dfrac{F}{dK}
\end{equation}

The above equation can be presented as

\begin{eqnarray*}
&&0.5\left( \left( q_{d}(t)-2q_{d}^{2}(t)+q_{d}^{3}(t)\right) d-\left(
q_{d}(t)-q_{d}^{2}(t)\right) \left( 1-\frac{n(t)}{K}\right) F\right) \\
&=&\dfrac{F}{dK}\left( \left( 0.5F+\Phi \right) \left( n(t)-\frac{n(t)^{2}}{%
K }\right) -n(t)\left[ \left( 1-2q_{d}(t)+q_{d}(t)^{2}\right) 0.5d+\Psi %
\right] \right)
\end{eqnarray*}
\bigskip

\begin{eqnarray*}
&&0.5\left( q_{d}(t)-2q_{d}^{2}(t)+q_{d}^{3}(t)\right) d-\left(
q_{d}(t)-q_{d}^{2}(t)\right) 0.5F+\left( q_{d}(t)-q_{d}^{2}(t)\right) \frac{
n(t)}{K}0.5F \\
&=&\dfrac{F}{dK}\left( \left( 0.5F+\Phi \right) n(t)-\left( 0.5F+\Phi
\right) \frac{n(t)^{2}}{K}-n(t)\left[ \left( 1-2q_{d}(t)+q_{d}(t)^{2}\right)
0.5d+\Psi \right] \right)
\end{eqnarray*}

\begin{eqnarray*}
&&0.5\left( q_{d}(t)-2q_{d}^{2}(t)+q_{d}^{3}(t)\right) d-\left(
q_{d}(t)-q_{d}^{2}(t)\right) 0.5F+\left( q_{d}(t)-q_{d}^{2}(t)\right) \frac{
n(t)}{K}0.5F \\
&=&\dfrac{F}{dK}\left( \left( 0.5F+\Phi -\left(
1-2q_{d}(t)+q_{d}(t)^{2}\right) 0.5d-\Psi \right) n(t)-\left( 0.5F+\Phi
\right) \frac{n(t)^{2}}{K}\right)
\end{eqnarray*}
\begin{equation*}
0.5\left( q_{d}(t)-2q_{d}^{2}(t)+q_{d}^{3}(t)\right) d-\left(
q_{d}(t)-q_{d}^{2}(t)\right) 0.5F+\left( q_{d}(t)-q_{d}^{2}(t)\right) \dfrac{
F}{K}0.5n(t)
\end{equation*}
\begin{eqnarray*}
&&0.5\left( q_{d}(t)-2q_{d}^{2}(t)+q_{d}^{3}(t)\right) d-\left(
q_{d}(t)-q_{d}^{2}(t)\right) 0.5F \\
&=&\left( 0.5F+\Phi -\left( 1-2q_{d}(t)+q_{d}(t)^{2}\right) 0.5d-\Psi
\right) \dfrac{F}{dK}n(t) \\
&&-\left( q_{d}(t)-q_{d}^{2}(t)\right) \dfrac{F}{K}0.5n(t)-\left( 0.5F+\Phi
\right) \dfrac{F}{dK^{2}}n(t)^{2}
\end{eqnarray*}

\begin{gather}
-\left( 0.5F+\Phi \right) \dfrac{F}{dK^{2}}n(t)^{2} \\
+\left[ \left( 0.5F+\Phi -\left( 1-2q_{d}(t)+q_{d}(t)^{2}\right) 0.5d-\Psi
\right) /d-\left( q_{d}(t)-q_{d}^{2}(t)\right) 0.5\right] \dfrac{F}{K}n(t) 
\notag \\
+0.5\left[ \left( q_{d}(t)-q_{d}^{2}(t)\right) F-\left(
q_{d}(t)-2q_{d}^{2}(t)+q_{d}^{3}(t)\right) d\right] =0  \notag
\end{gather}
$\bigskip $

Therefore, we have a quadratic equation with coefficients 
\begin{eqnarray*}
A_{q} &=&-\left( 0.5F+\Phi \right) \dfrac{F}{dK^{2}} \\
B_{q} &=&\left[ \left( 0.5F+\Phi -\left( 1-2q_{d}(t)+q_{d}(t)^{2}\right)
0.5d-\Psi \right) /d-\left( q_{d}(t)-q_{d}^{2}(t)\right) 0.5\right] \dfrac{F 
}{K} \\
C_{q} &=&0.5\left[ \left( q_{d}(t)-q_{d}^{2}(t)\right) F-\left(
q_{d}(t)-2q_{d}^{2}(t)+q_{d}^{3}(t)\right) d\right] ,
\end{eqnarray*}

and the subnullcline will be one of the roots of the standard form.
Numerical trajectories converge to 
\begin{equation*}
\mathcal{S}^{\varepsilon}_q(q_{d})=\frac{-B_{q}+\sqrt{B_{q}^{2}-4A_{q}C_{q}}}{2A_{q}}
\end{equation*}

\section{Appendix 4}\label{Appendix 4}

Now let us calculate subnullcline for the density nullcline. Density
nullcline can be presented as 
\begin{equation*}
n(q)=\left( 1-\dfrac{\left[ 1-2q_{d}(t)+q_{d}(t)^{2}\right] 0.5d+\Psi }{
0.5F+\Phi }\right) K,
\end{equation*}
and the derivative is 
\begin{eqnarray*}
\frac{dn}{dq} &=&\dfrac{\left[ 2-2q_{d}(t)\right] 0.5d}{0.5F+\Phi }K \\
&=&\dfrac{\left( 1-q_{d}(t)\right) dK}{0.5F+\Phi }
\end{eqnarray*}

Therefore 
\begin{equation*}
\frac{dq}{dn}=\dfrac{0.5F+\Phi }{\left( 1-q_{d}(t)\right) dK},
\end{equation*}

thus on the subnullcline will be satisfied equation for the equality of
slopes%
\begin{equation*}
\frac{g(x,y)}{f(x,y)}=\dfrac{0.5F+\Phi }{\left( 1-q_{d}(t)\right) dK},
\end{equation*}

which have form 
\begin{equation*}
\dfrac{0.5q_{d}(t)\left( 1-q_{d}(t)\right) \left( \left( 1-q_{d}(t)\right)
d-\left( 1-\frac{n(t)}{K}\right) F\right) }{n(t)\left( \left( 0.5F+\Phi
\right) \left( 1-\frac{n(t)}{K}\right) -\left[ \left( 1-q_{d}(t)\right)
^{2}0.5d+\Psi \right] \right) }=\dfrac{0.5F+\Phi }{\left( 1-q_{d}(t)\right)
dK}.
\end{equation*}
Let us transform the algebraic equation: 
\begin{equation*}
\dfrac{0.5q_{d}(t)\left( 1-q_{d}(t)\right) \left( \left( 1-q_{d}(t)\right)
d-\left( 1-\frac{n(t)}{K}\right) F\right) }{n(t)\left( \left( 0.5F+\Phi
\right) \left( 1-\frac{n(t)}{K}\right) -\left[ \left( 1-q_{d}(t)\right)
^{2}0.5d+\Psi \right] \right) }=\dfrac{0.5F+\Phi }{\left( 1-q_{d}(t)\right)
dK},
\end{equation*}
\begin{eqnarray*}
&&0.5q_{d}(t)\left( 1-q_{d}(t)\right) ^{2}\left( \left( 1-q_{d}(t)\right)
d-\left( 1-\frac{n(t)}{K}\right) F\right) dK \\
&=&n(t)\left( 0.5F+\Phi \right) \left( \left( 0.5F+\Phi \right) \left( 1- 
\frac{n(t)}{K}\right) -\left[ \left( 1-q_{d}(t)\right) ^{2}0.5d+\Psi \right]
\right) ,
\end{eqnarray*}
\begin{eqnarray*}
&&0.5q_{d}(t)\left( 1-q_{d}(t)\right) ^{3}d^{2}K-0.5q_{d}(t)\left(
1-q_{d}(t)\right) ^{2}\left( 1-\frac{n(t)}{K}\right) FdK \\
&=&n(t)\left( 0.5F+\Phi \right) ^{2}\left( 1-\frac{n(t)}{K}\right)
-n(t)\left( 0.5F+\Phi \right) \left[ \left( 1-q_{d}(t)\right) ^{2}0.5d+\Psi %
\right] ,
\end{eqnarray*}
\begin{eqnarray*}
&&0.5q_{d}(t)\left( 1-q_{d}(t)\right) ^{3}d^{2}K-0.5q_{d}(t)\left(
1-q_{d}(t)\right) ^{2}FdK+n(t)0.5q_{d}(t)\left( 1-q_{d}(t)\right) ^{2}Fd \\
&=&n(t)\left( 0.5F+\Phi \right) ^{2}-\left( 0.5F+\Phi \right) ^{2}\frac{
n(t)^{2}}{K}-n(t)\left( 0.5F+\Phi \right) \left[ \left( 1-q_{d}(t)\right)
^{2}0.5d+\Psi \right] ,
\end{eqnarray*}
\begin{eqnarray*}
&&0.5q_{d}(t)\left( 1-q_{d}(t)\right) ^{2}dK\left[ \left( 1-q_{d}(t)\right)
d-F\right] \\
&=&n(t)\left( 0.5F+\Phi \right) ^{2}-\left( 0.5F+\Phi \right) ^{2}\frac{
n(t)^{2}}{K} \\
&&-n(t)\left( 0.5F+\Phi \right) \left[ \left( 1-q_{d}(t)\right)
^{2}0.5d+\Psi \right] -n(t)0.5q_{d}(t)\left( 1-q_{d}(t)\right) ^{2}Fd,
\end{eqnarray*}
\begin{eqnarray*}
&&0.5q_{d}(t)\left( 1-q_{d}(t)\right) ^{2}dK\left[ \left( 1-q_{d}(t)\right)
d-F\right] \\
&=&n(t)\left[ \left( 0.5F+\Phi \right) ^{2}-\left( 0.5F+\Phi \right) \left[
\left( 1-q_{d}(t)\right) ^{2}0.5d+\Psi \right] -0.5q_{d}(t)\left(
1-q_{d}(t)\right) ^{2}Fd\right] \\
&&-\left( 0.5F+\Phi \right) ^{2}\frac{n(t)^{2}}{K},
\end{eqnarray*}

\begin{eqnarray*}
&&\dfrac{\left( 0.5F+\Phi \right) ^{2}}{K}n(t)^{2} \\
&&-\left[ \left( 0.5F+\Phi \right) ^{2}-\left( 0.5F+\Phi \right) \left[
\left( 1-q_{d}(t)\right) ^{2}0.5d+\Psi \right] -0.5q_{d}(t)\left(
1-q_{d}(t)\right) ^{2}Fd\right] n(t) \\
&&+0.5q_{d}(t)\left( 1-q_{d}(t)\right) ^{2}dK\left[ \left( 1-q_{d}(t)\right)
d-F\right] .
\end{eqnarray*}

Therefore, we have quadratic equation with coefficients 
\begin{eqnarray*}
A_{n} &=&\dfrac{\left( 0.5F+\Phi \right) ^{2}}{K} \\
B_{n} &=&-\left[ \left( 0.5F+\Phi \right) ^{2}-\left( 0.5F+\Phi \right) %
\left[ \left( 1-q_{d}(t)\right) ^{2}0.5d+\Psi \right] -0.5q_{d}(t)\left(
1-q_{d}(t)\right) ^{2}Fd\right] \\
C_{n} &=&0.5q_{d}(t)\left( 1-q_{d}(t)\right) ^{2}dK\left[ \left(
1-q_{d}(t\right) )d-F\right] ,
\end{eqnarray*}
and again, the subnullcline will be one of the roots of the standard form.
Numerical trajectories converge to 
\begin{equation*}
\mathcal{S}^{\varepsilon}_n(q_{d})=\frac{-B_{n}+\sqrt{B_{n}^{2}-4A_{n}C_{n}}}{2A_{n}}.
\end{equation*}

\section{Appendix 5}\label{Appendix 5}

\begin{figure}[htbp]
\centering

\begin{subfigure}{0.48\textwidth}
    \centering
    \includegraphics[width=\linewidth]{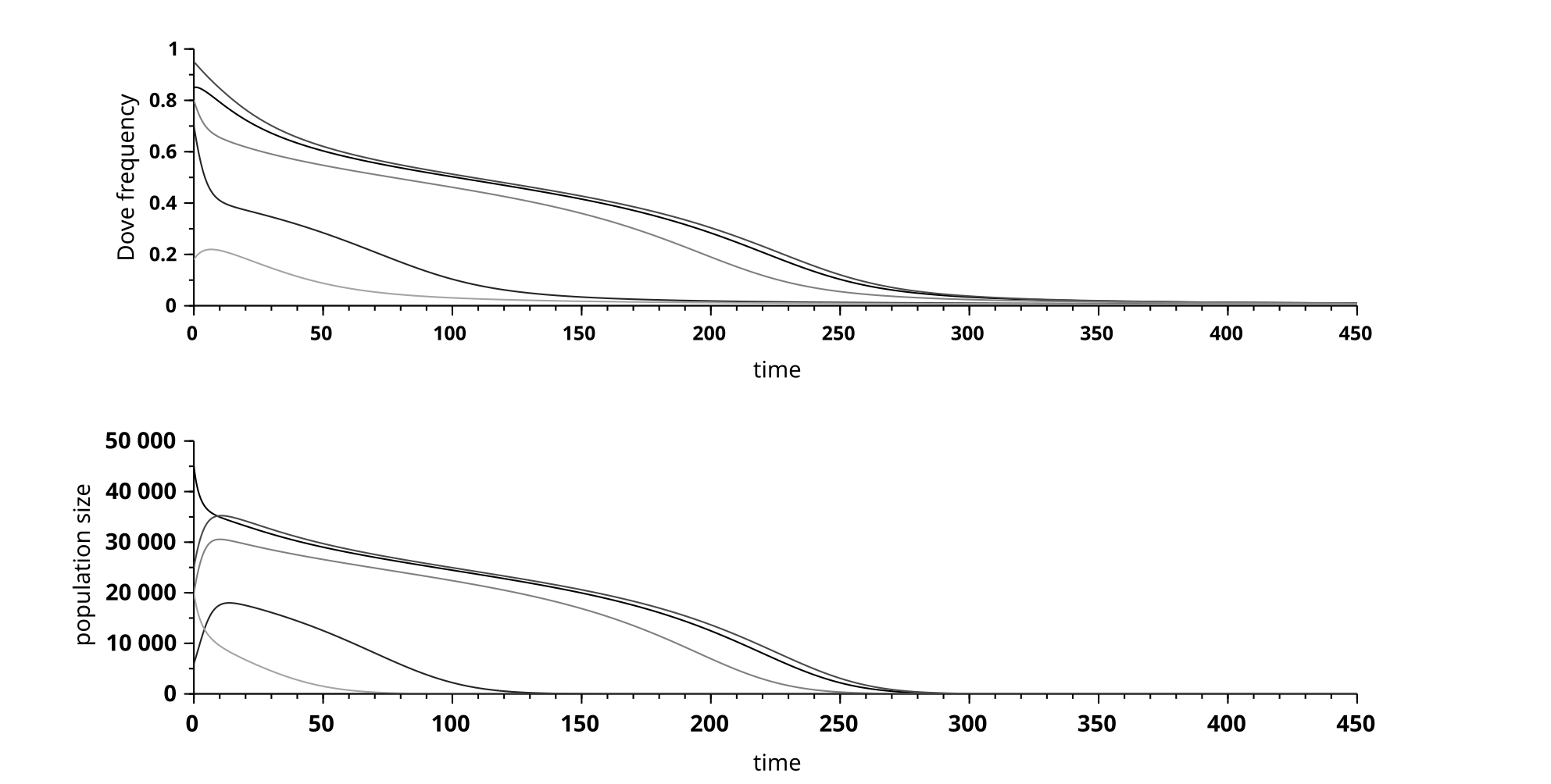}
    \caption{For $\Psi=0.01$, the duration of the long transient is significantly shorter.}
    \label{fig:WiderChan01Time}
\end{subfigure}
\hfill
\begin{subfigure}{0.48\textwidth}
    \centering
    \includegraphics[width=\linewidth]{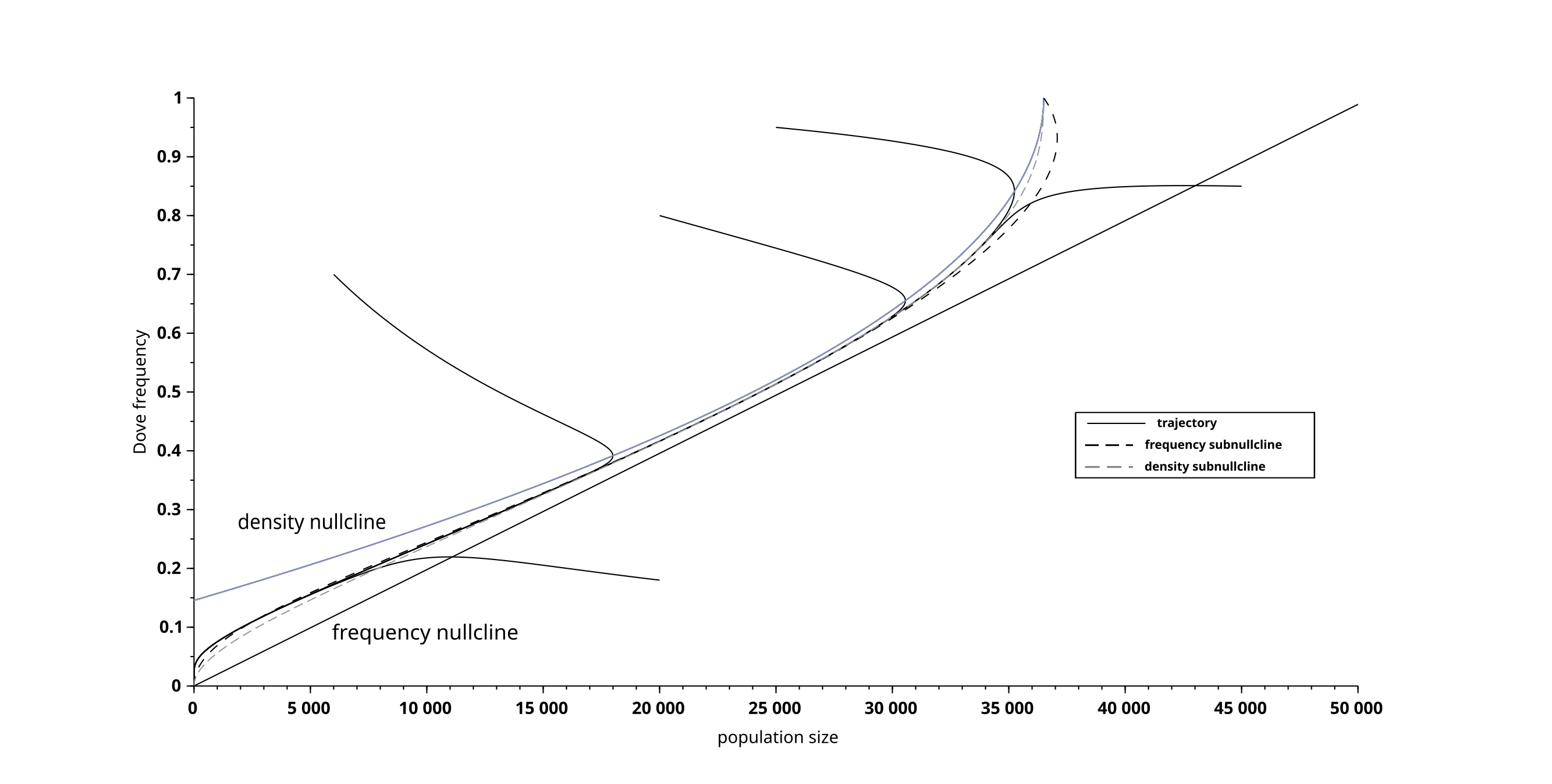}
    \caption{Phase portrait for the example in panel~\subref{fig:WiderChan01Time}. In this case, the channel between nullclines is clearly visible. However, subnullclines still provide a good approximation of the bundle of trajectories.}
    \label{fig:WiderChan01Phase}
\end{subfigure}

\vspace{0.4cm}

\begin{subfigure}{0.48\textwidth}
    \centering
    \includegraphics[width=\linewidth]{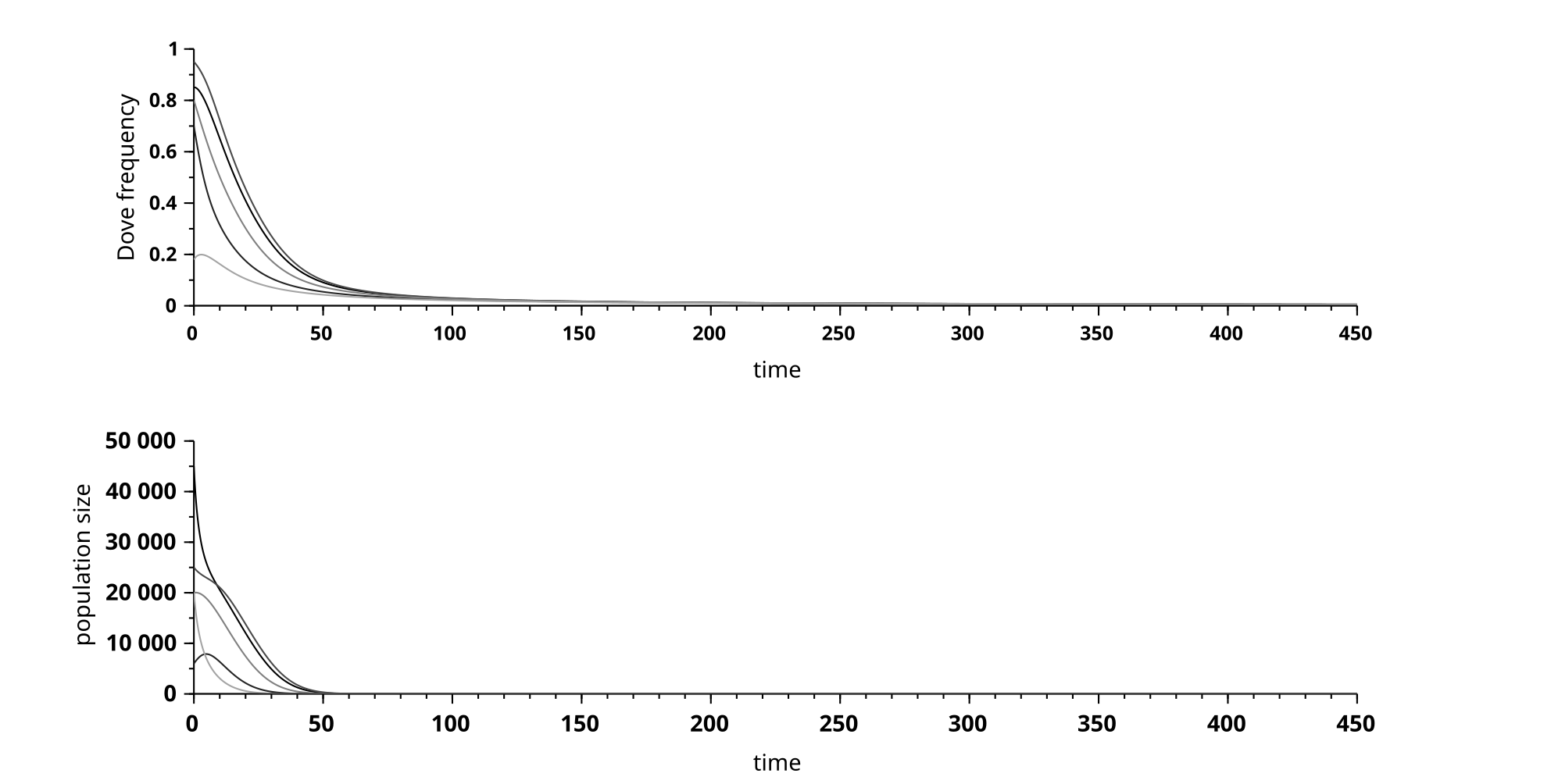}
    \caption{For $\Psi=0.15$, the long transient pattern disappears.}
    \label{fig:WiderChan15Time}
\end{subfigure}
\hfill
\begin{subfigure}{0.48\textwidth}
    \centering
    \includegraphics[width=\linewidth]{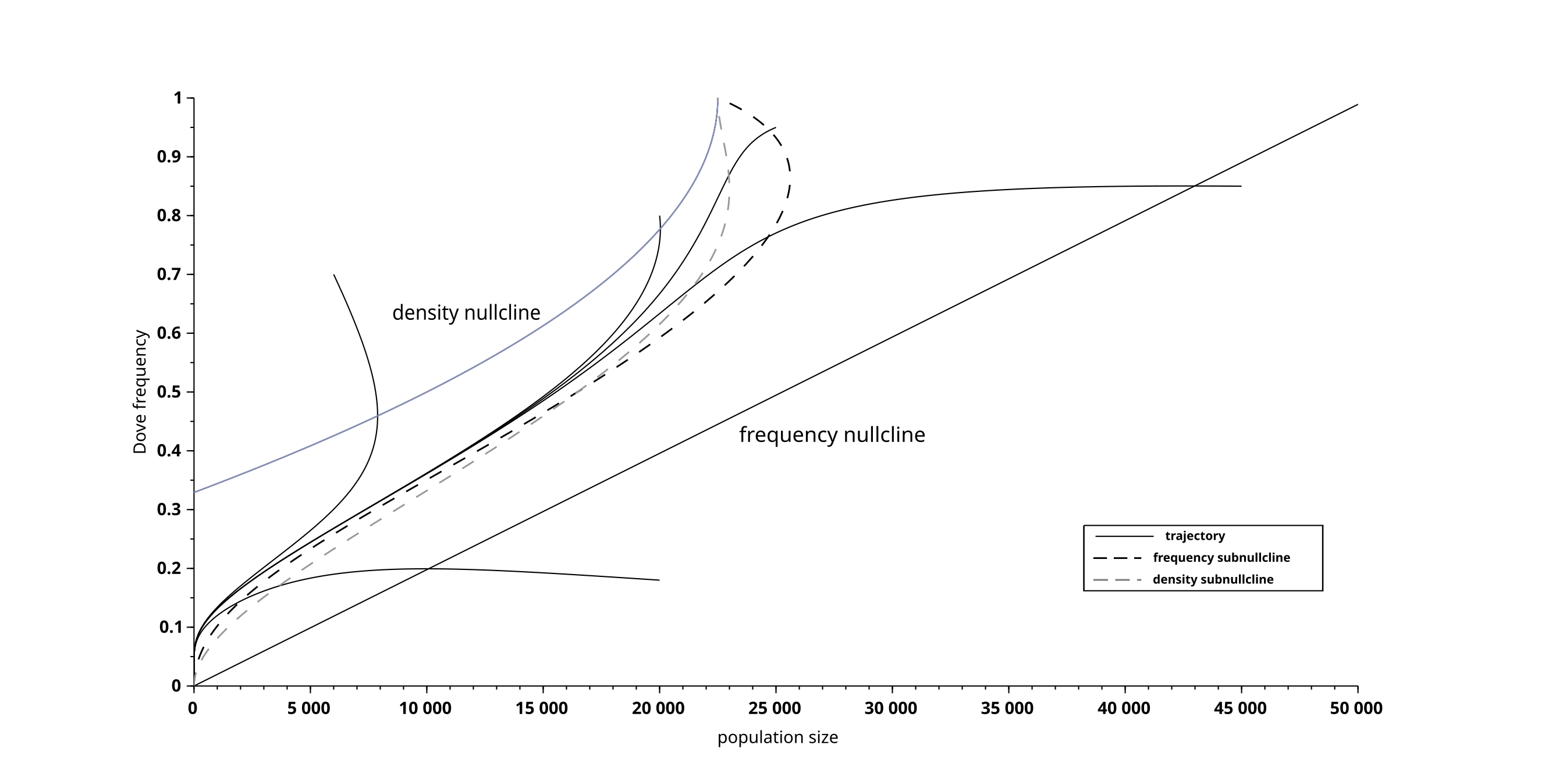}
    \caption{Phase portrait for the example in panel~\subref{fig:WiderChan15Time}. In this case, the channel between nullclines is very wide. Subnullclines still provide an acceptable approximation of the bundle of trajectories.}
    \label{fig:WiderChan15Phase}
\end{subfigure}

\vspace{0.4cm}

\begin{subfigure}{0.55\textwidth}
    \centering
    \includegraphics[width=10cm]{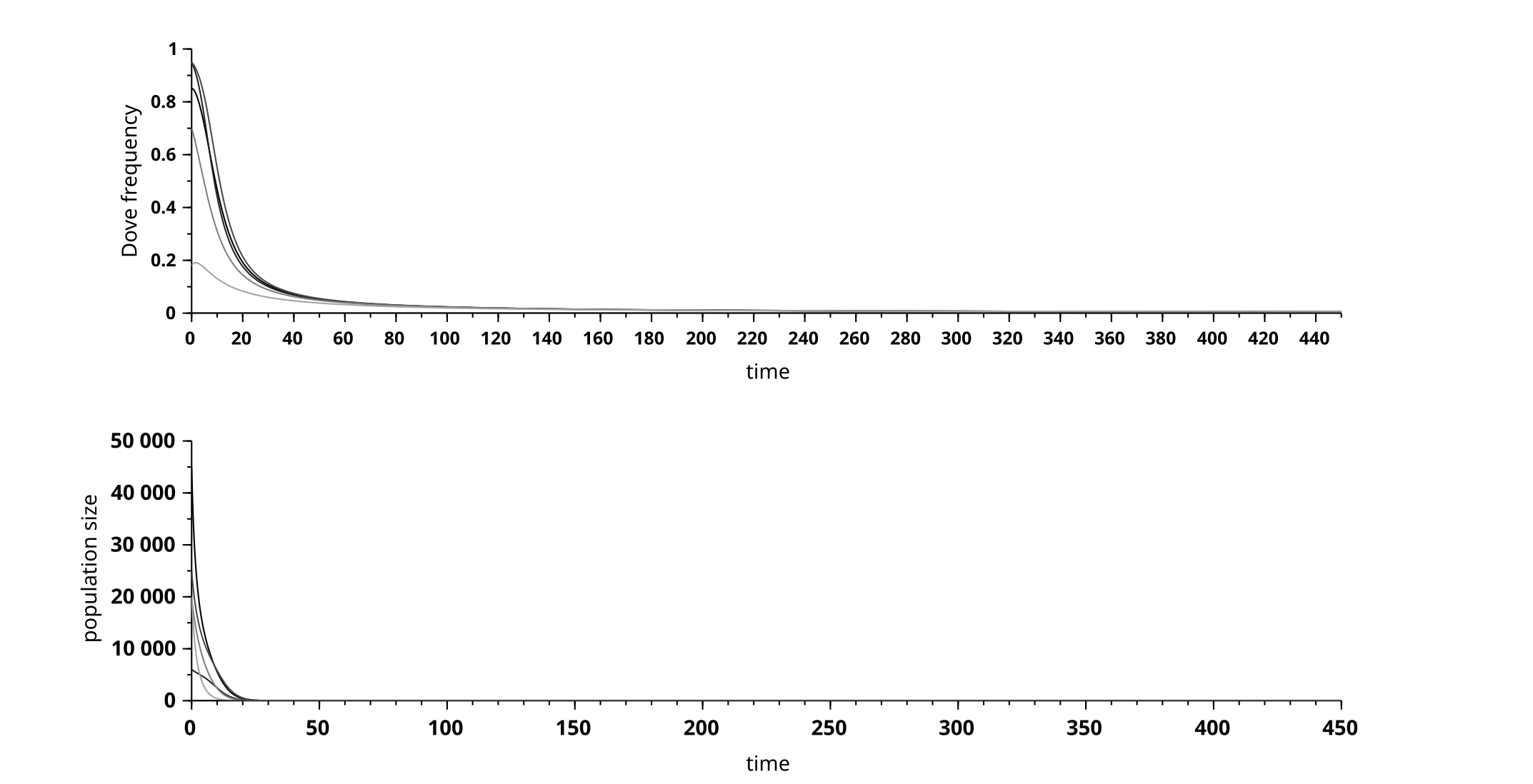}
    \caption{For $\Psi=0.37$ the long transient pattern is not present. }
    \label{FAIL-Time}
\end{subfigure}
\hfill
\begin{subfigure}{0.4\textwidth}
    \centering
    \includegraphics[width=\linewidth]{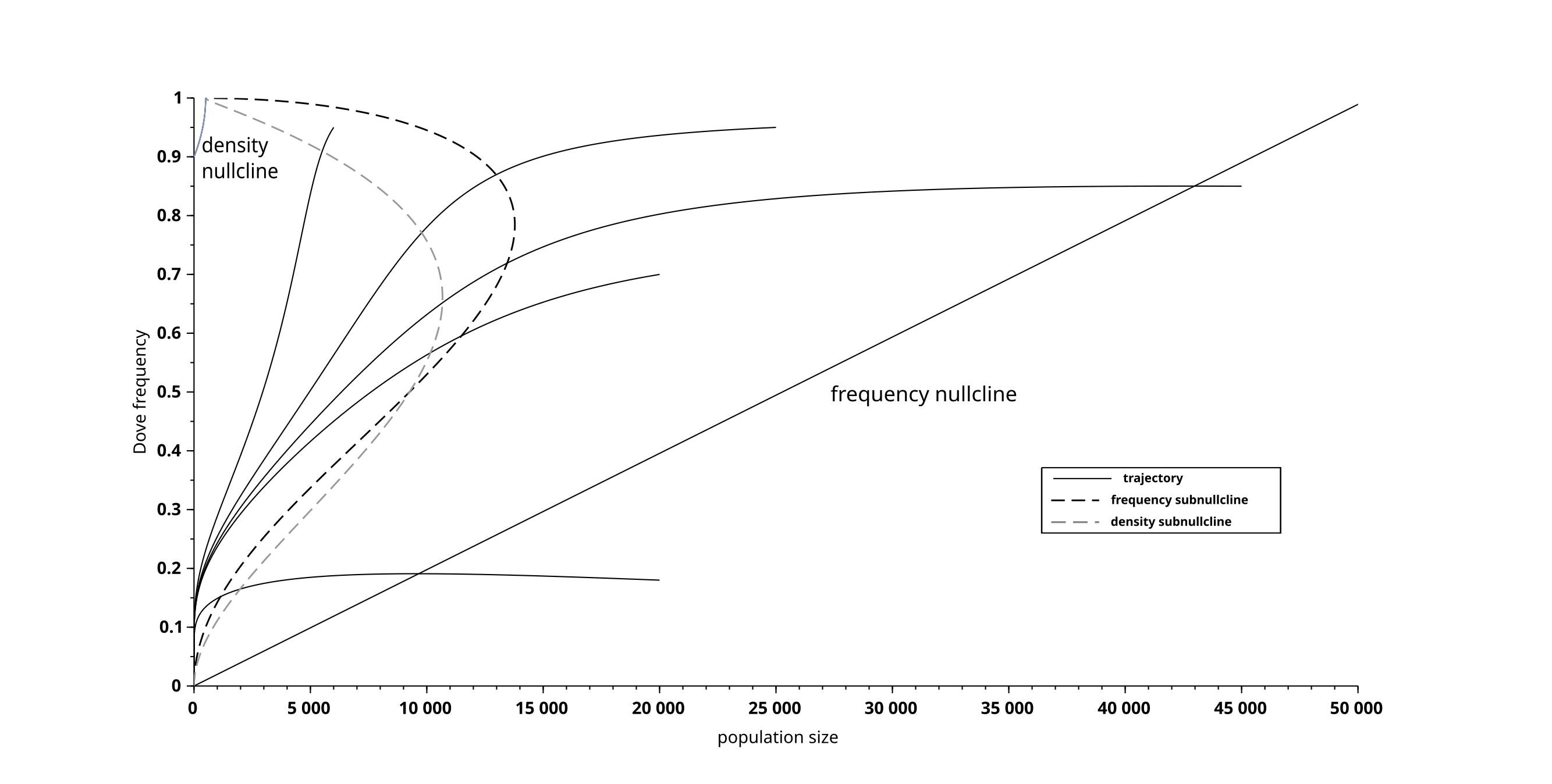}
    \caption{Phase portrait for the example from panel \subref{FAIL-Time}.
    Nullclines are far from each other. We observe that the accuracy of the
    approximation by the subnullclines deteriorates significntly. }
    \label{FAIL-Phase}
\end{subfigure}
\caption{Comparison of the system dynamics for wider channels between nullclines. Panels (a) and (b) show the time series and phase portrait for $\Psi=0.01$, panels (c) and (d) show the corresponding results for $\Psi=0.15$ and panels (e) and (f) show the corresponding results for $\Psi=0.37$.}
\label{fig:WiderChannelComparison}
\end{figure}

When the channel between the nullclines becomes wider, the duration of the long transients decreases. This is presented in Fig.~\ref{fig:WiderChannelComparison}, where panels~\subref{fig:WiderChan01Time} and~\subref{fig:WiderChan01Phase} show the case $\Psi=0.01$, while panels~\subref{fig:WiderChan15Time} and~\subref{fig:WiderChan15Phase} show the case $\Psi=0.15$. However, the subnullclines still provide an acceptable approximation. When the distance between the nullclines is large, the accuracy of the approximation based on subnullclines also deteriorates; see panels~\subref{FAIL-Time} and~\subref{FAIL-Phase} for the case $\Psi=0.37$.

\newpage

\bibliographystyle{unsrt}
\bibliography{References}

@article{doebeli,
  author  = {Doebeli, M. and Ispolatov, Y. and Simon, B.},
  title   = {Towards a mechanistic foundation of evolutionary theory},
  journal = {eLife},
  volume  = {6},
  pages   = {e23804},
  year    = {2017}
}

@article{argbr1,
  author  = {Argasinski, K. and Broom, M.},
  title   = {Ecological theatre and the evolutionary game: how environmental and demographic factors determine payoffs in evolutionary games},
  journal = {Journal of Mathematical Biology},
  volume  = {67},
  pages   = {935--962},
  year    = {2013}
}

@article{argbr2,
  author  = {Argasinski, K. and Broom, M.},
  title   = {Interaction rates, vital rates, background fitness and replicator dynamics: how to embed evolutionary game structure into realistic population dynamics},
  journal = {Theory in Biosciences},
  volume  = {137},
  pages   = {33--50},
  year    = {2018}
}

@article{argbr3,
  author  = {Argasinski, K. and Broom, M.},
  title   = {Evolutionary stability under limited population growth: Eco-evolutionary feedbacks and replicator dynamics},
  journal = {Ecological Complexity},
  volume  = {34},
  pages   = {198--212},
  year    = {2018}
}

@article{hui,
  author  = {Hui, C.},
  title   = {Carrying capacity, population equilibrium, and environment's maximal load},
  journal = {Ecological Modelling},
  volume  = {192},
  pages   = {317--320},
  year    = {2006}
}

@article{maynard1,
  author  = {Maynard Smith, J. and Price, G. R.},
  title   = {The logic of animal conflict},
  journal = {Nature},
  volume  = {246},
  pages   = {15},
  year    = {1973}
}

@book{maynard2,
  author    = {Maynard Smith, J.},
  title     = {Evolution and the Theory of Games},
  publisher = {Cambridge University Press},
  address   = {Cambridge},
  year      = {1982}
}

@book{hofsig1,
  author    = {Hofbauer, J. and Sigmund, K.},
  title     = {The Theory of Evolution and Dynamical Systems},
  series    = {London Mathematical Society Student Texts},
  volume    = {7},
  publisher = {Cambridge University Press},
  year      = {1988}
}

@book{hofsig2,
  author    = {Hofbauer, J. and Sigmund, K.},
  title     = {Evolutionary Games and Population Dynamics},
  publisher = {Cambridge University Press},
  address   = {Cambridge},
  year      = {1998}
}

@book{BroomRychtar,
  author    = {Broom, M. and Rycht{\'a}{\v{r}}, J.},
  title     = {Game-theoretical Models in Biology},
  publisher = {CRC Press},
  year      = {2022}
}

@book{Sinervo,
  author    = {Friedman, D. and Sinervo, B.},
  title     = {Evolutionary Games in Natural, Social, and Virtual Worlds},
  publisher = {Oxford University Press},
  year      = {2016}
}

@article{Hastings1,
  author  = {Hastings, A.},
  title   = {Transients: the key to long-term ecological understanding?},
  journal = {Trends in Ecology \& Evolution},
  volume  = {19},
  number  = {1},
  pages   = {39--45},
  year    = {2004}
}

@article{Hastings2,
  author  = {Hastings, A. and Abbott, K. C. and Cuddington, K. and Francis, T. and Gellner, G. and Lai, Y. C. and Zeeman, M. L.},
  title   = {Transient phenomena in ecology},
  journal = {Science},
  volume  = {361},
  number  = {6406},
  pages   = {eaat6412},
  year    = {2018}
}

@article{Morozov,
  author  = {Morozov, A. and Abbott, K. and Cuddington, K. and Francis, T. and Gellner, G. and Hastings, A. and Zeeman, M. L.},
  title   = {Long transients in ecology: Theory and applications},
  journal = {Physics of Life Reviews},
  volume  = {32},
  pages   = {1--40},
  year    = {2020}
}

@article{Morozov2,
  author  = {Morozov, A. and Feudel, U. and Hastings, A. and Abbott, K. C. and Cuddington, K. and Heggerud, C. M. and Petrovskii, S.},
  title   = {Long-living transients in ecological models: Recent progress, new challenges, and open questions},
  journal = {Physics of Life Reviews},
  volume  = {51},
  pages   = {423--441},
  year    = {2024}
}

@article{Koch2024,
  author  = {Koch, D. and Nandan, A. and Ramesan, G. and Tyukin, I. and Gorban, A. and Koseska, A.},
  title   = {Ghost channels and ghost cycles guiding long transients in dynamical systems},
  journal = {Physical Review Letters},
  volume  = {133},
  number  = {4},
  pages   = {047202},
  year    = {2024}
}

@article{ResHolling1,
  author  = {Holling, C. S.},
  title   = {Resilience and stability of ecological systems},
  journal = {Annual Review of Ecology and Systematics},
  volume  = {4},
  pages   = {1--23},
  year    = {1973}
}

@incollection{ResHolling2,
  author    = {Holling, C. S.},
  title     = {Engineering resilience versus ecological resilience},
  booktitle = {Engineering within Ecological Constraints},
  editor    = {Schultze, Peter},
  pages     = {31--43},
  year      = {1996}
}

@book{ResGund,
  editor    = {Gunderson, L. H. and Allen, C. R. and Holling, C. S.},
  title     = {Foundations of Ecological Resilience},
  publisher = {Island Press},
  year      = {2012}
}

@article{ResMeyer,
  author  = {Meyer, K.},
  title   = {A mathematical review of resilience in ecology},
  journal = {Natural Resource Modeling},
  volume  = {29},
  number  = {3},
  pages   = {339--352},
  year    = {2016}
}

@article{ResKrak,
  author  = {Krakovsk{\'a}, H. and Kuehn, C. and Longo, I. P.},
  title   = {Resilience of dynamical systems},
  journal = {European Journal of Applied Mathematics},
  volume  = {35},
  number  = {1},
  pages   = {155--200},
  year    = {2024}
}

@article{ResReed,
  author  = {Reed, J. M. and Wolfe, B. E. and Romero, L. M.},
  title   = {Is resilience a unifying concept for the biological sciences?},
  journal = {iScience},
  year    = {2024}
}

@article {Bedi2025,
	author = {Bedi, M. S. and Argasinski, K. and Broom, M.},
	title = {Eco-Evolutionary game dynamics with Three Strategies: how geometric analysis provides insights into stability},
	elocation-id = {2025.12.26.696515},
	year = {2025},
	doi = {10.64898/2025.12.26.696515},
	publisher = {Cold Spring Harbor Laboratory},
	URL = {https://www.biorxiv.org/content/early/2025/12/26/2025.12.26.696515},
	eprint = {https://www.biorxiv.org/content/early/2025/12/26/2025.12.26.696515.full.pdf},
	journal = {bioRxiv}
}

\end{document}